%% file: archiveX_ms.tex
\documentclass[12pt,preprint]{aastex}

\usepackage{url}
\usepackage{graphics,graphicx}
\usepackage{color}

\newcommand{\ltsimeq}{\raisebox{-0.6ex}{$\,\stackrel
        {\raisebox{-.2ex}{$\textstyle <$}}{\sim}\,$}}
\newcommand{\gtsimeq}{\raisebox{-0.6ex}{$\,\stackrel
        {\raisebox{-.2ex}{$\textstyle >$}}{\sim}\,$}}

\def\H0{{\rm ~km~s^{-1}~Mpc^{-1}}}

\shorttitle{IR Spectrophotometry of C/2012 K1 (Pan-STARRS)}
\shortauthors{Woodward et al.}

\slugcomment{ApJ Accepted 2015.July.22}
\received{}
\revised{}

\begin{document}\sloppy

\title{SOFIA Infrared Spectrophotometry of \\
Comet C/2012 K1 (Pan-STARRS)}


\author{Charles E. Woodward\altaffilmark{1},
Michael S.~P. Kelley\altaffilmark{2},
David E. Harker\altaffilmark{3},
Erin L. Ryan\altaffilmark{2}, \\
Diane H. Wooden\altaffilmark{4},
Michael L. Sitko\altaffilmark{5},
Ray W. Russell\altaffilmark{6},\\
William T. Reach\altaffilmark{7},
Imke de Pater\altaffilmark{8},
Ludmilla Kolokolova\altaffilmark{2},
Robert D. Gehrz\altaffilmark{9}
}


\altaffiltext{1}{Minnesota Institute for Astrophysics, University
of Minnesota, 116 Church St, SE Minneapolis, MN 55455, USA; chickw024@gmail.com}

\altaffiltext{2}{University of Maryland, Department of Astronomy,
College Park, MD 20742-2421, USA}

\altaffiltext{3}{University of California, San Diego, Center for 
Astrophysics \& Space Sciences, 9500 Gilman Dr. Dept 0424, La Jolla, 
CA 92093-0424, USA}

\altaffiltext{4}{NASA Ames Research Center, MS 245-3, Moffett Field,
CA 94035-0001, USA}

\altaffiltext{5}{Department of Physics, University of Cincinnati, 
Cincinnati, OH 45221, USA}

\altaffiltext{6}{The Aerospace Corporation, Los Angeles, CA 90009, USA}

\altaffiltext{7}{USRA-SOFIA Science Center, NASA Ames Research Center, Moffett Field, 
CA 94035, USA}

\altaffiltext{8}{Astronomy Department, 601 Campbell Hall, University of 
California, Berkeley, CA 94720, USA}

\altaffiltext{9}{Minnesota Institute for Astrophysics, University
of Minnesota, 116 Church St, SE Minneapolis, MN 55455, USA}


\begin{abstract}
We present pre-perihelion infrared 8 to 31~\micron{} spectrophotometric 
and imaging observations of comet C/2012 K1 (Pan-STARRS), a dynamically new 
Oort Cloud comet, conducted with NASA's Stratospheric 
Observatory for Infrared Astronomy (SOFIA) facility (+FORCAST) in 2014 June. 
As a ``new'' comet (first inner solar system 
passage), the coma grain population may be 
extremely pristine, unencumbered by a rime and insufficiently 
irradiated by the Sun to carbonize its surface organics. The comet 
exhibited a weak 10~\micron{} silicate feature $\simeq 1.18
\pm 0.03$ above the underlying best-fit $215.32 \pm 0.95$~K 
continuum blackbody.
Thermal modeling of the observed spectral energy distribution 
indicates that the coma grains are fractally solid with a porosity 
factor $D = 3$ and the peak in the grain size distribution, 
$a_{peak} = 0.6$~\micron{}, large. The sub-micron coma grains are dominated 
by amorphous carbon, with a silicate-to-carbon ratio of 
$0.80^{ +0.25}_{- 0.20}$. The silicate crystalline mass fraction 
is $0.20^{ +0.30}_{ -0.10}$, similar to
with other dynamically new comets exhibiting weak 10~\micron{} silicate 
features. The bolometric dust albedo of the coma dust 
is $0.14 \pm 0.01$ at a phase angle of 34.76\degr, and the average
dust production rate, corrected to zero phase, at the epoch of 
our observations was $Af\rho \simeq 5340$~cm. \\
\end{abstract}


\section{INTRODUCTION}

Solar System formation was an engine that simultaneously preserves and 
transforms interstellar medium (ISM) ices, organics, and dust grains 
into cometesimals, planetesimals and, ultimately, planets. 
Observing and modeling the properties of small, primitive bodies in
the solar system whose origins lie beyond the water frost line
($>5$~AU) provides critical insight into the
formation of Solar System solids and establishes
observation constraints for planetary system formation invoking 
migration -- the `Grand Tack' epoch \citep{walsh2011}, followed 
by the `Nice Model' events \citep{levison2009,gomes2005}.
The characteristics of comet dust can provide evidence to validate 
the new, emerging picture of small body populations -- including 
comet families -- resulting from planetary migration in the 
early Solar System.

Inside cometary nuclei, the bulk of the dust likely has been preserved
since formation of the nucleus. Comet grains (and ices) also trace 
the pre-accretion history of comet
materials extant in the outer disk. Comet dust composition can be studied
via \textit{Stardust} samples, selected collections of
Interplanetary Dust Particles (IDPs), and in situ
analysis in comet flyby and/or rendezvous missions.
 Dust species that are best explained
as products of aqueous alteration (e.g., magnitite, cubanite, possibly
pentlandite) are rare \citep{stodolna2012, berger2011, zolensky2008}
and corresponding altered silicates (e.g., phyllosilicates, smectite) are 
missing suggesting that aqueous alteration in cometary nuclei is 
limited, is not well represented 
in the \textit{Stardust} samples, or that these minerals
have exogenous origins \citep{brownlee2014}. Thus, the bulk of comet
grain properties including dust size, porosity, and composition
relate to grain formation, radial mixing, and particle agglomeration in
the proto-solar disk \citep[for an extensive review see][]{brownlee2014}.
However, opportunities to study actual samples of cometary dust 
are rare, motivating the need for telescopic remote
sensing observations of dust whenever apparitions are
accessible from terrestrial observatories.

In this paper we report our pre-perihelion (TP = 2014 Aug 27.65~UT)
infrared 8 to 31 micron spectrophotometric 
observations of comet C/2012 K1 (Pan-STARRS), a dynamically new
\citep[see][for a definition based on orbital elements]{oort1950}
Oort Cloud comet -- $(1/a_{org}) = 42.9 \times 10^{-6}$~AU$^{-1}$
\citep{williams2015} -- conducted with NASA's Stratospheric Observatory 
for Infrared Astronomy (SOFIA) facility during a series of four 
flights over the period from 2014 June 04 to 13~UT. Contemporaneous optical
imaging observations are also presented.

\section{OBSERVATIONS}

\subsection{Ground-based Optical Imaging}\label{obs-bok}

Comet C/2012 K1 (Pan-STARRS) was observed on 2012 June 01.22~UT and again on
June 04.24~UT with the 2.3-m Bok Telescope at the Kitt Peak
National Observatory.  The comet was at heliocentric distance ($r_{h}$) of
1.74~AU and 1.71~AU, a geocentric distance ($\Delta$) of 1.66~AU and 1.69~AU,
a phase angle of $34.62^{\circ}$ and $34.76^{\circ}$, for
each date respectively.

The images were obtained with the 90Prime camera \citep{williams04}, a
prime focus imager built for the Bok Telescope. At the time of
observation, the 90Prime camera utilized a thinned back-illuminated CCD
detector with 4064 $\times$ 4064 pixels with a pixel size of
$15.0~\micron$. At prime focus the camera pixel scale is
$0.45^{\prime\prime}$ which yields a field of view of 30.5 $\times$ 30.5
square-arcmin. The instrument was equipped with Cousins/Bessel system
broadband $V$ and $R$ filters. Multiple exposures (23 images 
in $R$ band and 9 images in $V$ band of 30 seconds each) were obtained 
of the nucleus and coma of the comet with the telescope tracking at the
non-sidereal rate corresponding to the predicted motion of the comet
provided by JPL Horizons\footnote{http://ssd.jpl.nasa.gov/horizons.cgi}
in an airmass range of 1.40 to 1.74. 

\begin{figure}
\plotone{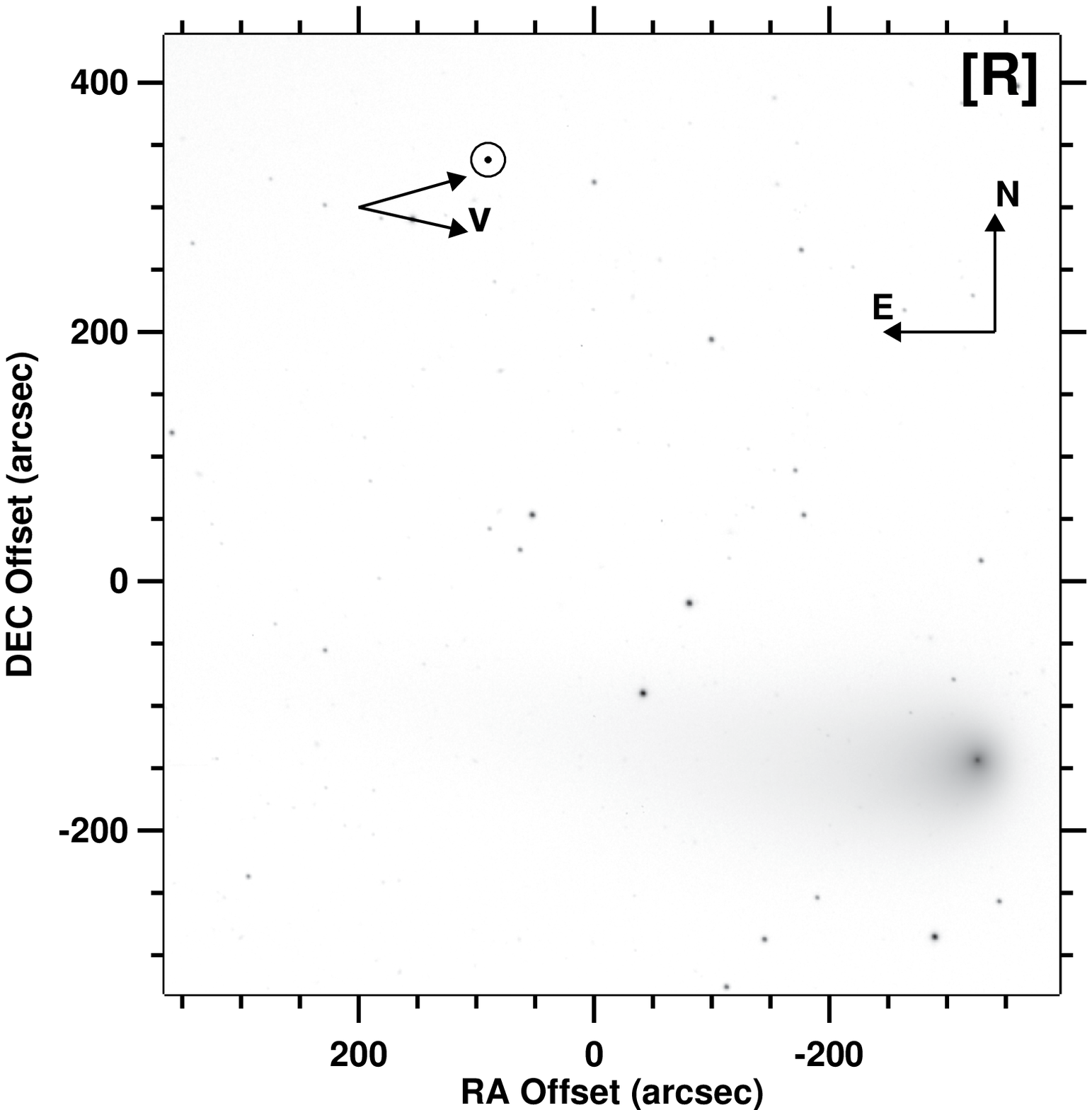}
\caption{The Bok $R$-band optical images of comet C/2012 K1 (Pan-STARRS) 
obtained on 2014 June 04.24~UT, with logarithmic greyscale color map. The 
vector indicating the direction of the comet's motion and the 
vector indicating the direction
toward the Sun are also provided.
\label{fig:bokVandR}}
\end{figure}

All images were corrected for
overscan, bias and flat-fielding with standard IRAF\footnote{IRAF is
distributed by the National Optical Astronomy Observatory, which is
operated by the Association of Universities for Research in Astronomy
(AURA) under cooperative agreement with the National Science Foundation.}
routines.  The data was photometrically calibrated using eight field stars
of various spectral types with known $V$ and $R$ magnitudes
selected from the Naval Observatory Merged Astrometric
Dataset (NOMAD) catalog \citep{zacharias2004} on the same
CCD amplifier as the comet. The standard deviation of the 
photometric $V$ and $R$ zero points derived from the average of the 
field stars is of order 1\% and no color
corrections for spectral type were applied.
The average nightly seeing was $\sim 2.2^{\prime\prime}$ in both bands.
A single 30~sec exposure in the $R$ band obtained on
2014 June 04.24~UT is shown in Fig.~\ref{fig:bokVandR}.

\subsection{SOFIA}\label{obs-sofia}

Mid-infrared (mid-IR) spectrophotometric observations of 
comet C/2012 K1 (Pan-STARRS) were 
obtained using the Faint Object InfraRed CAmera for the SOFIA 
Telescope \cite[FORCAST;][]{herter2012} mounted at the Nasmyth focus 
of the 2.5-m telescope of the SOFIA 
Observatory \citep{young2012, gehrz2009}. 
The data were acquired over a series of four flights, originating from
Palmdale, CA at altitudes of $\simeq 11.89$~km in 2014 June, that
were conducted as part of our SOFIA Cycle 2 programs to observe comets
(P.I. Woodward, AOR\_IDs 01\_001 and 02\_0002).
Details of all SOFIA observations and the orbital parameters of comet
C/2012 K1 (Pan-STARRS) at those epochs are summarized in
Table~\ref{tab:sobstab_tab}.

FORCAST is a dual-channel mid-IR imager and grism spectrometer 
operating from 5 to 40~\micron. Light is fed to two $256 \times 256$ pixel 
blocked-impurity-band (BIB) arrays, each with a plate scale of 
$0.768^{\prime\prime}$ per pixel and a distortion corrected field of 
view of $3.2^{\prime} \times 3.4^{\prime}$. The Short Wavelength Camera 
(SWC) covers the spectral region from 5 to 25~\micron{}, while the 
Long Wavelength Camera (LWC) operates at wavelengths from 25 to 40~\micron.
Imaging data can be acquired in either dual channel mode (with some
loss of throughput due to the dichroic) or single channel mode.

Imaging observations of C/2012 K1 (Pan-STARRS) in three filters were
conducted on the first flight series, prior to three flights dedicated
to spectroscopy. Spectroscopic observations of the 
comet used two grisms, one in the
SWC (G111) and one in the LWC (G227), and the instrument was 
configured using a long-slit ($4.7^{\prime\prime} \times
191^{\prime\prime}$) which yields a spectral 
resolution $R = \lambda/\Delta\lambda \sim$ 140-300. The 
comet was imaged in the SWC using the F197 filter
to position the target in the slit. Both imaging and spectroscopic data
were obtained using a 2-point chop/nod in the Nod-Match-Chop (C2N) mode 
with 45$^{\prime\prime}$ chop and 90$^{\prime\prime}$ nod amplitudes 
at angles of 30$^{\circ}$/210$^{\circ}$ in the equatorial reference frame. 

All FORCAST raw image data products were processed using the FORCAST\_REDUX Data
Pipeline, v1.0.1beta \citep[cf.,][]{clarke2014}, which employed 
the reduction packages FORCAST\_FSPEXTOOL, v1.1.0,  
and FORCAST\_DRIP, v1.1.0. Processing of the raw spectroscopic 
data was performed 
using the same packages, with the exception of FORCAST\_DRIP, which utilized 
v1.0.4. Details of the FORCAST\_REDUX Data Pipeline can be found in the Guest 
Investigator Handbook for FORCAST Data Products, 
Rev.~B\footnote{https://www.sofia.usra.edu/Science/DataProducts/\\ 
 FORCAST\_GI\_Handbook\_RevA1.pdf}

\subsection{SOFIA Imagery and Photometry}

Aperture photometry of the SOFIA image data of comet 
C/2012 K1 (Pan-STARRS) was performed on the Level 3 
pipeline coadded (*.COA) data products
using the Aperture Photometry Tool \citep[APT v2.4.7;][]{laher2012}. At all
FORCAST filter wavelengths, the comet exhibited extended emission beyond the
PSF of point sources observed with FORCAST under optimal telescope jitter
performance.\footnote{see http://www.sofia.usra.edu/Science/\\ 
ObserversHandbook/FORCAST.html \S5.1.2}
The photometry was therefore conducted using a circular aperture
centroided on the photocenter of the comet nucleus. We used an
aperture of radius 13 pixels, corresponding to 
9.984$^{\prime\prime}$, with a 
background aperture annulus of inner radius 30 pixels 
(23.58$^{\prime\prime}$) and outer radius of 60 pixels
(47.16$^{\prime\prime}$). This aperture, which is $\simeq 3 \times$ 
the nominal point-source FWHM, encompassed the majority of 
the emission of the comet and coma. Sky-annulus
median subtraction \citep[ATP Model B as described in][]{laher2012} 
was used in the computation of the source intensity.
The systematic source intensity uncertainty was computed using a depth of 
coverage value equivalent to the number of coadded image frames.
The dominant source of overall uncertainty in the image photometry
were image gradients due to imperfect atmospheric background subtraction.
The calibration factors (and associated uncertainties) applied to the
resultant aperture sums were included in the Level 3 
data distribution and were derived from the weighted average 
of 3 calibrator observations of $\beta$~And 
(2 each) and $\alpha$~Boo (1 each). The resultant SOFIA photometry 
is presented in Table~\ref{tab:simage_phot_tab}.

Due to turbulence, telescope jitter, and differing chop-nod 
patterns, i.e., the chopping difference between beams and the nodding of
the entire telescope field-of-view 
\citep[for a for a discussion and illustration
of this standard infrared observing technique
with SOFIA -- see][]{temi2014,young2012} executed
in flight, the multi-filter imagery data could not be used
to generate color temperature maps due to the unstable PSF. 

\begin{figure}
\plotone{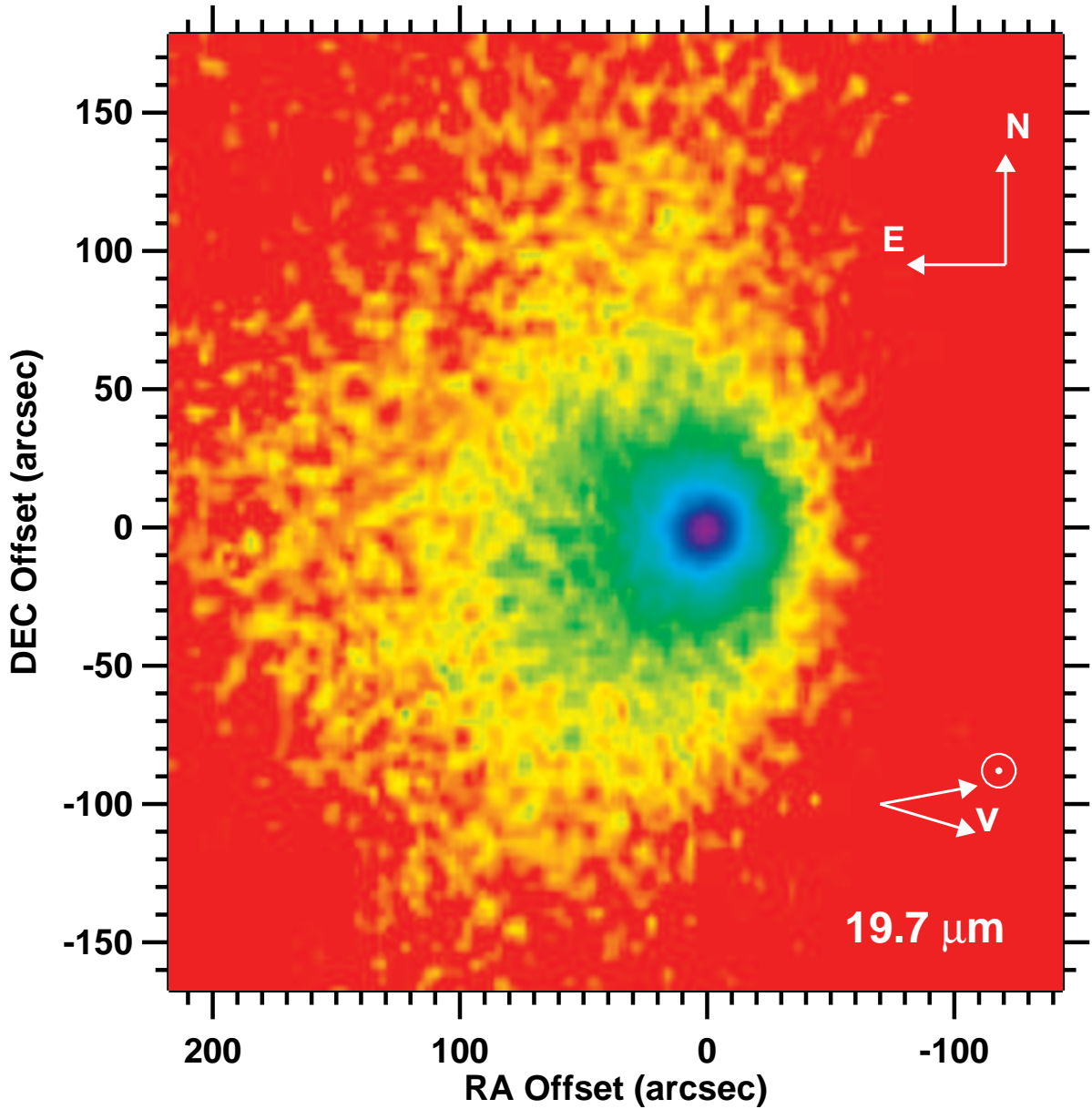}
\caption{The SOFIA 19.7~\micron{} image of comet C/2012 K1 (Pan-STARRS) obtained
on 2014 June 13.17~UT, with logarithmic color map. The vector indicating
the direction of the comet's motion and the vector indicating the direction
toward the Sun are also provided.
\label{fig:s195_image}}
\end{figure}

During flights primarily devoted to obtaining grism data (\S\ref{sec:ssp}),
images of the comet where obtained through a single filter at 19.7~micron.
Figure~\ref{fig:s195_image} shows the 19.7~\micron{} surface 
brightness distribution of comet C/2012 K1 (Pan-STARRS) observed 
on 2014 June 13.17~UT. The nucleus is unresolved and azimuthally symmetric 
with a radial profile FWHM of $\sim 1.01^{\prime\prime}$ and the
coma is extended and diffuse. Low surface brightness emission extends
in a vector direction commensurate with that expected for a dust tail.

\subsection{SOFIA Spectra}\label{sec:ssp}

\begin{figure}
\plotone{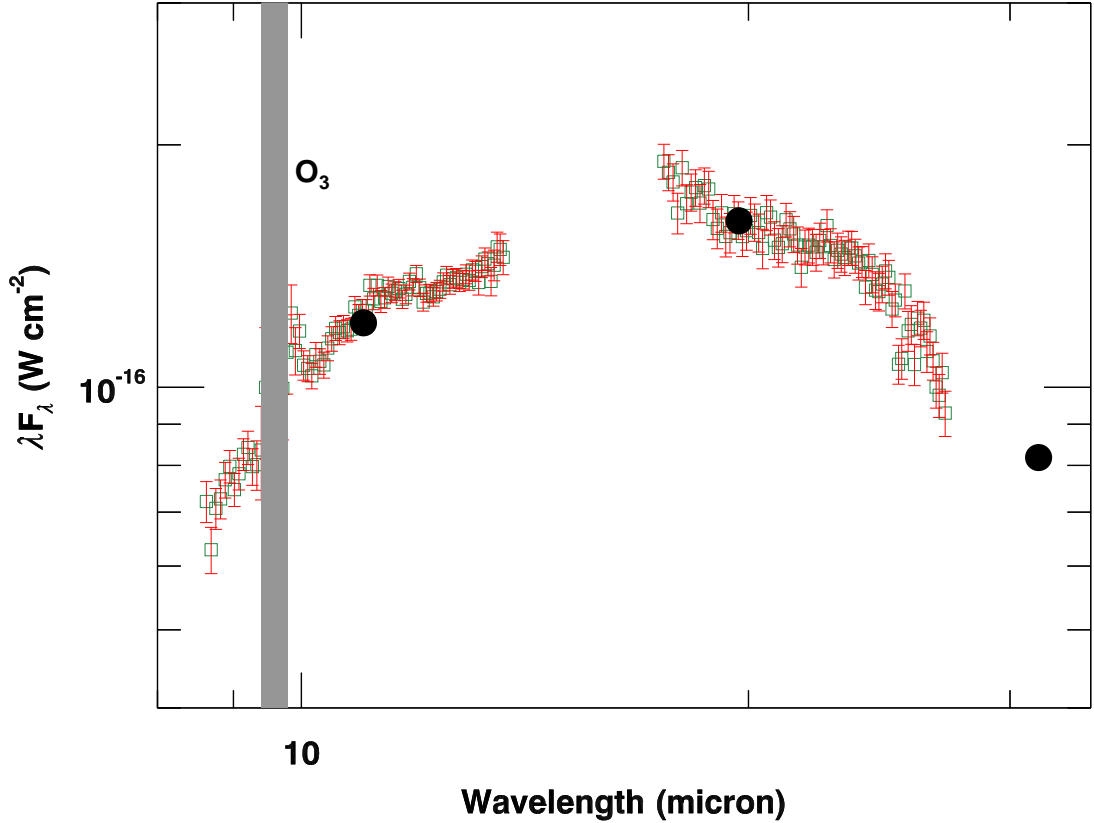}
\caption{The SOFIA grism observations of comet C/2012 K1 (Pan-STARRS). 
Grism observations from three flight series where combined to 
create a single average SED, and a three-point 
unweighted rectangular smoothing function was applied to the
composite spectra to improve the signal-to-noise. The filled circles 
are the broadband photometric observations obtained on 2015 June 04 UT 
when each of the filters were observed on a single flight. The 
photometry is scaled by a factor of 0.75 (which is of the order 
of absolute photometric calibration uncertainty of FORCAST) to match the flux
density of the grism spectra and demonstrate that the shape of 
the photometric SEDs mimics that of the spectroscopic spectral 
segments. The vertical grey bar indicates the region of 
terrestrial ozone absorption. 
\label{fig:sofia_grating}}
\end{figure}

Three temporally distinct spectra of comet C/2012 K1 (Pan-STARRS) 
were obtained in both grism over a series of flight sequences spanning 6 days 
(Table~\ref{tab:sobstab_tab}). Many comets exhibit temporal variability 
in the infrared over periods of hours \citep[e.g.,][]{wooden2004}
to days at relatively similar heliocentric distances due to coma jets related 
to nucleus activity and/or nucleus rotation 
period \citep[e.g.,][]{keller2007,gehrz1995}
that produces observable changes in the observed SEDs. Inter-comparison of 
each SED over this period showed no substantial changes in
overall continuum flux densities nor spectral features to within the 
uncertainty per spectral resolution element. In addition, the 
19.7~\micron{} aperture photometry suggests also that the level of 
coma emission did not markedly change (cf., Table~\ref{tab:simage_phot_tab}). 
Apparently, comet C/2012 K1 (Pan-STARRS) was 
fairly quiescent in its infrared behavior during this epoch given
our signal-to-noise ratio and aperture size (12,300~km radius). Thus, 
the three independent spectra were summed together in pipeline 
processing to produce an average spectral energy distribution (SED). A 3-point 
unweighted rectangular smoothing function was applied to this average SED 
to increase the point-to-point signal-to-noise ratio of the data product 
used in our thermal model spectral decomposition analysis. The 
calibrated data products do exhibit a few artifacts near the edges
of the 17--27~\micron{} spectral order where a few data points deviate upwards
(near 17~\micron) or downwards (near 27~\micron) from the apparent spectral 
trend. The spectra of comet C/2012 K1 (Pan-STARRS) are presented 
in Fig.~\ref{fig:sofia_grating}.

\section{RESULTS}

Taxonomically comet C/2012 K1 (Pan-STARRS)
is a member of the dynamical comet family denoted as nearly isotropic
comets (NICs), also commonly referred to as Oort Cloud comets
\citep[cf.,][]{dones2004}. The interior composition of the ecliptic 
comets (ECs) and the NICs likely are preserved during their residence in 
the Scattered Disk and the Oort Cloud, but their surfaces are 
subject to various processing effects. Modeling the coma dust 
properties provides insight into the 
origin and evolution of dynamic comet families.

\subsection{Thermal Modeling of the Coma SED}\label{sec:sofia_ans}

Thermal modeling of the observed thermal infrared SED of 
comets obtained using remote sensing techniques enables
derivation of coma dust grain properties. In particular, 
SOFIA (+FORCAST) provides spectroscopic coverage with the 
G111 grism to the region 9--12~\micron{} which contains features 
from amorphous and crystalline silicates (e.g., 11.2~\micron) 
and organic species (e.g., PAHs). The G227 grism spans
17.6--27.7~\micron{}, encompassing discrete resonances from crystalline 
silicates as well as spectral signatures from
carbonates and phyllosilicates, putatively argued to be
extant in comets \citep{lisse2006}. The SED slope at long thermal 
(\gtsimeq 15~\micron) wavelengths provides constraints on the 
abundance of the larger grain population in the coma.
Observations in these spectral regimes are key to
ascertaining the origins of silicates within
the solar protoplanetary disk, and placing early solar disk evolution
within the context of other circumstellar disks observed today
through comparison to model and laboratory
data \citep[cf.,][]{lindsay2013,koike2010}.

Modeling the mid-IR SED of C/2012 K1 (Pan-STARRS) yields
estimates of the coma grain properties. We constrain the grain parameters
by chi-squared fitting thermal emission models to the observed
spectrum. The grain parameters included in the modeling are
size distributions (n(a)\, da), porosity,
the crystalline mass fraction (i.e., the fraction of the
coma silicate grains that are crystalline), and
relative material abundances.  The dust temperature is calculated
assuming thermal equilibrium of the grains; wherein the composition
(mineralogy), size, and heliocentric distance determine the
temperature of the grains. 

Comet grains are dominated in composition by a handful of 
silicate-type materials \citep{hannerzol2010,wooden2008}: Mg-Fe olivine- 
and pyroxene-types in amorphous (glassy) forms 
and their crystalline Mg-end-members forsterite (Mg$_{2}$SiO$_{4}$) 
and enstatite (MgSiO$_{3}$). Cometary aggregates also contain organics 
\citep{sandford2006} or amorphous-carbon-like materials 
\citep{matrajt2008,formenkova1999}
that may be the glue that holds the amorphous and crystalline materials 
together \citep{flynn2013,cieslasanford2012}.
Our model \citep[][and references therein]{harker2002}
uses five materials: amorphous olivine and amorphous pyroxene with 
broad 10, 18,  and 20~\micron{} emission features, amorphous carbon 
with featureless emission, and crystalline olivine (Mg-rich) and 
orthopryoxene with narrow peaks. Broad and narrow resonances near 10 and
20~\micron{} are modeled by warm
chondritic (50\% Fe; 50\% Mg) amorphous silicates (i.e., glasses)
and strong 11.25, 19.5, and 24~\micron{} narrow features from
cooler Mg-rich crystalline silicate materials. 

The amorphous carbon component in our dust model is
representative of several key dust species -- e.g., elemental carbon
dust \citep{formenkova1994}, an organic component with
C=C bonds, identified by XANES spectra near 285~eV that can
include amorphous carbon \citep{flynn2013,flynn2003,wirick2009},
and possibly other carbonaceous grains; however,
overall model results do not depend on this degeneracy. We do not
specifically include Fe-Ni sulfides (such as pyrrhotite or troilite)
in our models nor carbonates or phyllosilicate-rich 
materials. The latter materials have not been detected in track analysis 
of \textit{Stardust} samples \citep{nakamura2011,wooden2008,zolensky2008} 
nor are they unequivocally evident in remote sensing
data \citep{bursentova2012,cew2007}. Phyllosilicates, specifically
smectites including montmorillonite, chlorite, and serpentine,
have 18-23~\micron{ } resonances that worsen spectral fitting of comet
C/1995 O1 (Hale-Bopp) \citep{wooden1999}. Hybrid IDPs may contain up to
10\% smectite \citep{nakamura2011}. Smectitie is spectrally
distinguishable from amorphous anhydrous olivine-type and 
amorphous pyroxene materials \citep{nakamura2011,wooden1999}, yet it is 
not required for spectral decomposition.

While FeS-type grains are 
present in IDPs \citep{bradleydai2000}, meteoritics samples, and comets grains, 
such as Wild 2 \citep{heck2012,zolensky2008,velbel2007}, our SOFIA spectra
(Fig.~\ref{fig:sofia_grating}) do not exhibit the broad
23~\micron{} spectral features often associated with fine-grained FeS
\citep{bursentova2012,min2005,hony2002,keller2002}. Larger FeS
particles would be spectrally indistinguishable from larger amorphous
carbon particles at mid- to far-IR wavelengths, yet robust optical
constants spanning visible through the far-IR are lacking for
FeS due to measurement challenges of an inherently extremely absorbing
material (L. Keller, private comm.). Thus, thermal modeling of FeS gains is uncertain.
\textit{Stardust} samples appear to be richer in FeS and poorer in carbonaceous
matter \citep{joswiak2012}, so there is no basis as yet to make an
assumption about the relative abundance of FeS and amorphous 
carbon-like materials in comet comae. Hence we presume
that the majority of absorbing materials in cometary dust 
re-radiating the observed infrared SED is dominated by olivine, pyroxene, and 
carbonaceous (amorphous carbon-like) 
materials \citep{brownlee2014,wooden2008,zolensky2008}.  This presumption provides
a foundation for comparing compositional similarities and diversities of comet
dust composition derived from thermal models.

\begin{figure}
\plotone{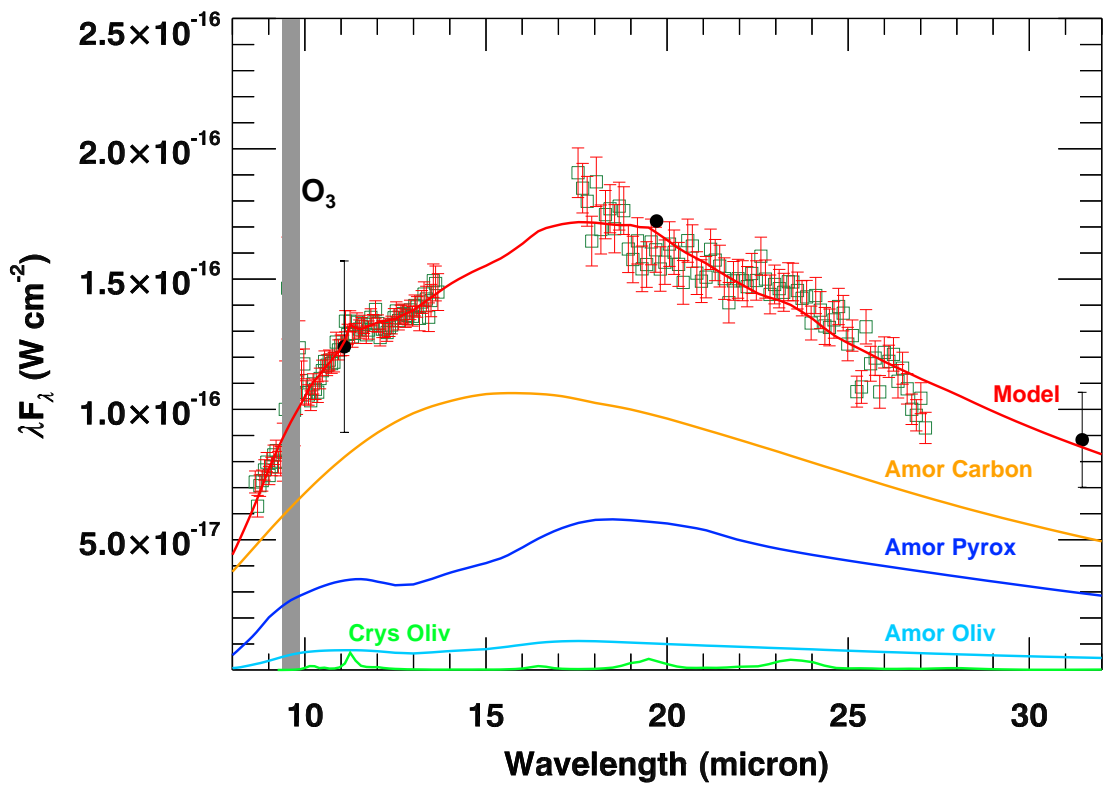}
\caption{The spectral decomposition of the SOFIA grism observations 
(see Fig.~\ref{fig:sofia_grating}) of comet C/2012 K1 (Pan-STARRS) 
derived from thermal modeling. The filled circles are the 
broadband photometric observations obtained on 2015 June 04 UT 
when each of the filters were observed on a single flight. The filter
photometry is scaled by a factor of 
0.75 (which is of the order of absolute photometric calibration 
uncertainty of FORCAST) to match the flux density of the grism spectra 
and demonstrate that the shape of the photometric SEDs mimics that
of the spectroscopic spectral segments. The vertical grey bar indicates 
the region of terrestrial ozone absorption.
\label{fig:sofia_model}}
\end{figure}

The best-fit chi-square model results are summarized in
Table~\ref{tab:bfmods_tab}. The model fit to the observed
grism spectra with the corresponding spectral decomposition
of grain components is presented in Fig.~\ref{fig:sofia_model}.
Mineralogically, the grains in the coma of C/2012 K1 (Pan-STARRS)
are dominated by amorphous materials, especially carbon.
Our models produce a Hanner (modified-power law) 
differential grain size distribution
(HGSD)\footnote{This power law (in grain radii, $a$) is defined as 
$n(a)\, da \equiv (1 - a_{o}/a)^{M} (a_{o}/a)^{N}$;
where $a_{o} = 0.1$~\micron{} and $a_{peak} = a_{o}(N+M)/N$.}
peaking with grains of radii $a_{peak} = 0.6$~\micron, indicating 
relative moderately larger
grains are present, and the grain power-law slope $N = 3.4$. In
a HGSD the small radii grains at the peak of the grain size distribution
dominate the surface area and the flux density.

Grains in comets are likely fractal porous 
aggregates \cite{schulz2015}. The grain
porosity ($P$ versus the dust radius $a$), parameterized by $D$, is 
defined as $P = (a/0.1~\mu\rm{m})^{(D-3)}$ with $ D = 3$ for
solid and $D = 2.5$ for highly porous grains \citep{cew2011}.
Grains in the coma of C/2012 K1 (Pan-STARRS) also are solid (the fractal
porosity parameter $D = 3.0$). Solid grains are not
unusual, 65\% of the \textit{Stardust} tracks are carrot-shaped from solid terminal
particles \citep{horz2006}. The sub-micron sized silicate-to-carbon ratio
derived from our models is $0.80^{+ 0.25}_{- 0.20}$. The uncertainty in the
parameters derived from the thermal models are at the 95\% confidence level.

\subsection{The 10~\micron{} Silicate Emission Feature}\label{sec:10sef-disc}

\begin{figure}
\plotone{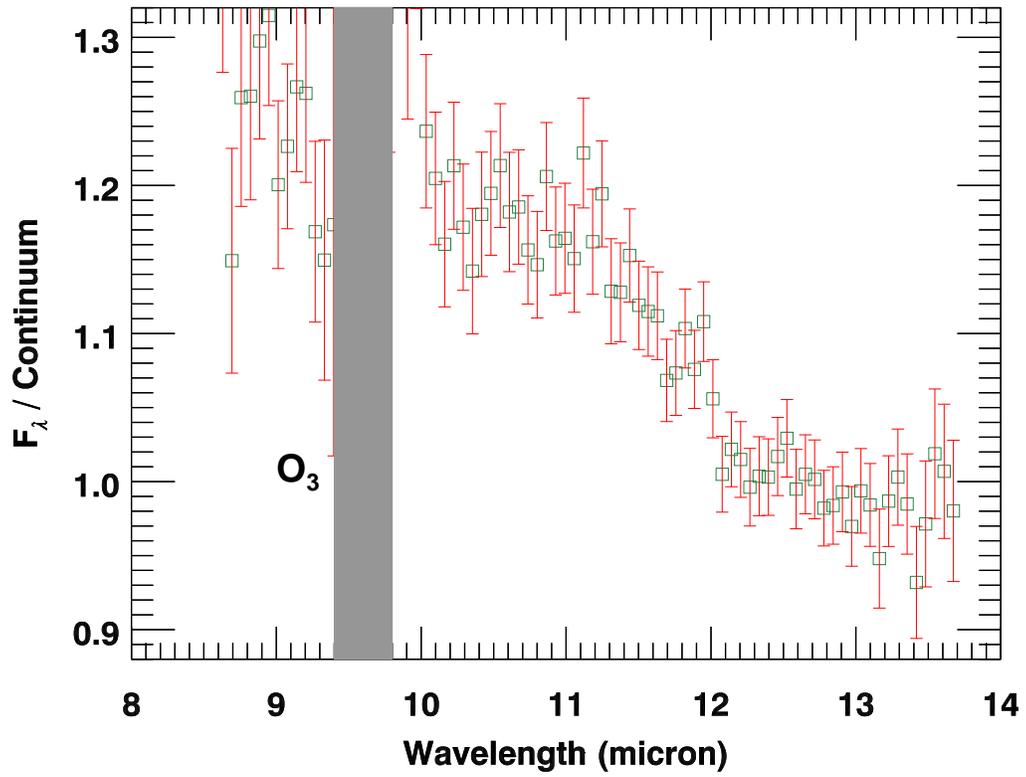}
\caption{The observed SOFIA grism flux density of
comet C/2012 K1 (Pan-STARRS) near the 10~\micron{} silicate emission
feature divided by a $\simeq 215$~K blackbody continuum 
($F_{\lambda}/F_{\lambda,T}$) to highlight the details of the 
10~\micron{} silicate feature. The vertical grey bar indicates the region of
terrestrial ozone absorption.
\label{fig:fbyfctbb}}
\end{figure}

The 10~\micron{} silicate feature in comet C/2012 K1 (Pan-STARRS) is
quite weak compared to comets like C/1995 O1 (Hale-Bopp)
or 17P/Holmes \citep[e.g.,][]{wooden1999, watanabe2009}. 
Following \citep{sitko2004}, at 10.5~\micron{} we find the
silicate emission (defined as $[F_{10}/F^{BB}_{continuum}]$)
is $1.18 \pm 0.03$ above a blackbody curve fit to the observed grism
spectra continua longwards of 12.5~\micron. The best-fit blackbody
is $T_{bb} = 215.32 \pm 0.95$~K (using Gaussian weighted errors)
and color excess, defined as T$_{bb}$(fit)/(278K $r_{h}^{-0.5}$)
is $= 0.992 \pm 0.004$. The normalized ($F_{\lambda}/F_{\lambda,T}$) 
SED in the region near the silicate feature at 10~\micron{} is presented in
Fig.~\ref{fig:fbyfctbb}. Typically data near
8~\micron{} are used to establish the blue-continua
($\lambda\lambda 7.7-8.4$~\micron). Our estimate of the local 
10~\micron{} continua may yield slightly lower temperatures
than an estimate that included 8~\micron{} photometry.

The large value of $a_{peak}$ inferred from the
thermal modeling of the observed
SED of comet C/2012 K1 (Pan-STARRS) in 2014 June is
commensurate with the weak 10~\micron{} silicate
feature. Smaller grains ($a_{peak} \ltsimeq 0.3$~\micron)
produce higher contrast silicate features. Grains of greater porosity also
produce higher contrast silicate features in the 10~\micron{} band.
Long period NICs have `typical' HGSD slopes of $3.4 \ltsimeq N \ltsimeq 3.7$
and silicate-to-amorphous carbon ratios $\gg 1$.
The grain size distribution slope of C/2012 K1 (Pan-STARRS), $N = 3.4$,
is not atypical. The preponderance of larger sub-micron
grains ($a_{peak} = 0.6$~\micron) in the coma of comet C/2012 K1
(Pan-STARRS) results in cooler radiating dust that
contributes to the `continuum' under the 10~\micron{} silicate feature and
to the far-infrared flux density (see Fig.~\ref{fig:sofia_model}). The
sub-micron mass fraction is dominated by amorphous carbon grains.
Amorphous carbon has a featureless emission spectrum that extends
through the 10~\micron{} region, so
low amorphous silicate-to-carbon ratio also can weaken the silicate feature
strength \citep{wooden2008, wooden2004}.

\subsection{The Silicate Crystalline Mass Fraction}\label{sec:fcryst-disc}

The mass fraction of silicate sub-micron grains that are crystalline
in comet comae is a keystone for models of early planet-forming processes 
\citep{bv2002,ciesla2007,ha2010}. This fraction is defined as

\begin{equation}
f_{cryst}^{silicates} \equiv \sum_{x=1}^{n} \frac{m_{cryst,x}}{(m_{cryst,x} + 
m_{amorphous,x}^{silicates})}
\label{eqn:fcseqn}
\end{equation}

\noindent where m$_{x}$ is the mass of species $x$. Crystalline 
species in comet grains provide a record of the 
high temperature process that formed dust in 
the inner disk of the solar system and the large scale mixing that 
transported these hot nebular products to the cold comet forming 
zones. Crystals, their composition \citep[e.g.,][]{wooden2008}
and shape \citep{lindsay2013} trace inner-solar disk
conditions \citep[e.g.,][]{ogliore2011} and offer a view into the 
earliest planet-forming processes that occurred in our early Solar System.

Crystals from the inner disk were transported out to the
comet-forming regime and mixed with ``amorphous'' 
silicates \citep[cf.][]{ciesla2011}.
The ``amorphous'' silicates  are thought to be outer
disk materials that probably were inherited from the ISM
\citep{brownlee2014,watson2009,kemper2004,kemper2005,lidraine2001} in the
infall phase of the disk. They have non-stoichiometric compositions
\citep[GEMS-like,][and references therein]{matsuno2012,bradleydai2004,bradley1999}
that include the compositional ranges of olivine 
[(Mg$_{y}\,$,\,Fe$_{(1-y)}$)$_{2}$~SiO$_{4}$], 
with y $\approx$ 0.5 for amorphous olivine, and
[(Mg$_{x}\,$,\,Fe$_{(1-x)}$)~SiO$_{3}$]
with x $\approx$ 0.5 for amorphous pyroxene-type materials.
Crystals are identified by narrow
IR emission features (e.g., 11.2, 19, 23.5, 27.5, 33~\micron)
superposed on an underlying thermal continuum in remote sensing
spectra.  Crystalline silicates have been 
detected using remote sensing techniques
in the dust comae of all comet classes including C/1995 O1
(Hale-Bopp) \citep{wooden1999, harker2002,harker2004a}
the Deep Impact coma of
9P/Tempel 1 \citep{harker2005,lisse2006,harker2007},
the fragmentation outburst of 17P/Holmes \citep{reach2010},
and several other comets \citep{mskdw2009,cew2011}.
Amorphous silicates are also detected in these comets as well.
Crystalline silicates are found in
abundance in the \textit{Stardust} samples of 81P/Wild 2; however,
the amorphous grains are difficult to identify \citep{ishii2008}
and may be limited to the smallest dust grains \citep{brownlee2014}.

Crystalline silicates are rare in the ISM; however, they account 
for \ltsimeq 2.2\% of the total silicate 
component in the direction of the Galactic 
Center \citep{kemper2004,kemper2005} and \ltsimeq 5\% along other
lines-of-sight \citep{lidraine2001}. Solar
System crystalline silicates detected in comets 
must be formed in the early stages of
our disk's evolution \citep{brownlee2014}. Crystalline silicates 
require T \gtsimeq 1000K to form through either gas phase condensation
or annealing of amorphous (glassy) silicate grains
\citep{wooden2008,wooden2005,davoisne2006,henning2003,fabian2000}
implying that the crystalline silicates must have been processed
in the disk near the young Sun or in shocks out to a
maximum distance of 3 to 5 AU \citep{dehsjd2002,wg2008}. Post-formation,
they were transported radially outward into the comet-formation
zones \citep{cm2007} -- a process that is apparently
ubiquitous in observations of external protoplanetary
disks \cite{olofsson2010}. Glassy silicate spherules (GEMS) and 
crystals are seen in aggregates in cometary IDPs. Large
`terminal particle' crystals and sub-micron crystals (crystallites) are 
components of aggregate grains captured in \textit{Stardust} samples 
\citep{brownlee2012,brownlee2006,zolensky2008,zolensky2006}.

Thus to first order, the diversity of comet
dust properties reflects the temporal and radial gradients in our
Solar System's early history and similarities and differences in dust
characteristics, including $f_{cryst}$,  may provide 
observational tests of of planetary migration models within 
the early solar system during the epoch of planet formation
that resulted in a variety of small body dynamical populations.
We find a that the silicate crystalline mass 
fraction in comet C/2012 K1 (Pan-STARRS) is
$f_{cryst} = 0.20^{+ 0.30}_{- 0.10}$. This range is 
similar to that found for comet C/20007 N3 (Lulin), 
$f_{cryst} = 0.48 \pm 0.06$ \citep[derived from the
mass fractions presented in Table~3 of][and Eqn.~\ref{eqn:fcseqn}
of \S\ref{sec:fcryst-disc}]{cew2011}
which also exhibited a weak 10~\micron{} silicate 
emission feature.

\subsection{EC and NIC Dust Characteristics}\label{sec:ecnic-disc}

As a result of giant planet migration, some comet nuclei 
were dynamically scattered into the Oort cloud to be 
exposed to the Galactic environment, whereas 
those bodies comprising 
the bulk of the ECs population have nuclei exposed and processed (at
depths ranging from  mm to few cm) by solar insolation, space weathering, 
and heliocentric activity variations (sublimation of CO, CO$_{2}$;
crystallization of water and other ices) which 
affects materials lofted into the comae. 
Although the interior compositions of ECs
and NICs likely are preserved, their surfaces have differing processing
histories. 

Typically, active comets (arising from a population of
NICs dynamically derived from the Oort Cloud and moving on 
long-period orbits) exhibit high contrast 10~$\micron${} silicate
features. In contrast, short-period ECs
(i.e., Jupiter-family comets) have, on average, lower 10~\micron{}
silicate features strengths \citep{sitko2004}, and are thought
to have lower activities \citep[cf., the active area and active
fraction measurements of][]{ahearn1995}.  For decades, the low-activity 
of ECs has been attributed to the accumulation of a rime of 
insulating larger grains that were launched on non-escape
orbits \citep{jewitt2007}. Thermal models that fit observed infrared spectra
of comets reveal that high contrast silicate features arise from comae 
having a preponderance of sub-micron grains \citep{harker2002,hanner1994}. 
Comae without these sub-micron grains have weaker silicate features.
In individual comets, variations in the silicate
feature strength have been seen on short time
scales corresponding to the aperture-crossing times of
jets or coma features \citep{wooden2004,harker2005,harker2007,gicquel2012}.
These variations are best explained by changes in the differential
grain size (n(a)\, da) or fluctuations in the 
silicate-to-carbon grain ratio.

Differences between EC and NIC coma grain populations may 
arise from the surface layers EC nuclei being 
``processed'' or weathered \citep[e.g.,][]{lijy2015}. Processing 
of ECs surfaces may result from their frequent perihelion passages 
that decreases surface volatiles and small grains and leads to the
creation of rimes and dust mantles. Evidence suggesting
such processing occurs over millennia may be found in the analysis of material
excavated from comet 9P/Tempel~1 by \textit{Deep Impact}: the dust grains
in the ejecta were smaller than those in the ambient coma \citep{harker2007}
and the immediate comet surface contained a layer of carbon rich grains
\citep{sugita2005} and a dust mantle comprised of compact 20~\micron{}-sized
dust aggregates \citep{kobayashi2013}. However, this conjecture is not 
definitive as it unknown whether or not the impact location reflects the 
global surface dust properties of the nucleus. In ECs, the 
coma 10~\micron{} silicate feature strengths are low \citep{mskdw2009}
and the dust production rates are modest. 
However, when EC nuclei have either fragmented 
\citep[i.e., 73P/SW3,][]{harker2011, sitko2011}, 
explosively released materials from subsurface cavities 
\citep[i.e., 17P/Holmes,][]{reach2010}, or 
have had subsurface materials excavated from depth 
\citep[i.e., the 9P/Tempel~1 \textit{Deep Impact} encounter,][]{harker2005}
the IR SEDs exhibit 10~\micron{} relatively strong silicate feature 
emission ($\gtsimeq 1.2$) arising from a population of 
sub-micron size silicate grain species. Whether or not the strong
10~\micron{} silicate features arise from the release of sub-micron sized grains 
or the disruption of loose aggregates of fine particles (e.g., through 
gas-pressure disruption or impact fragmentation)  is not known. Indeed the 
silicate feature in 9P/Tempel~1 changed from a EC-like spectra to NIC-like 
spectra immediately after \textit{Deep Impact} event, returning to an EC-like 
state several tens of hours later \citep[cf.,][]{harker2005}.

NICs are canonically considered to be more pristine with
higher surface volatile abundance \citep[cf.,][]{wooden2008} -- the
effects of dwell time in the Galactic environment being more benign. Also,
NICs are often considered a homologous population
lacking significant nucleus evolution. Inner solar system 
apparitions of these comets frequently result in brilliant comae, 
with large dust production rates and pronounced silicate 
feature emission at IR wavelengths. It is not entirely clear 
whether the highly active nuclear regions of NICs can spawn small 
sub-micron grains responsible for the silicate feature emission, either 
by heritage or by fragmentation induced within the gas acceleration 
zone.  However, whether comet evolution, such as processing in 
the Galactic environment, can be ignored when
comparing the Oort cloud comet dust composition (including that 
expressed in $f_{cryst}$)
is an open question. 

Table~\ref{tab:fcryst_tab} present estimates of 
$f_{cryst}$ and select characteristics
of the dust derived from thermal modeling of the mid-IR SEDs for 
a set of well-studied Oort cloud and ``disrupted'' Jupiter-family comets.
The crystalline silicate fraction ranges appreciably, from 
$\sim10$\% to $\sim80$\%. The compositional similarity 
suggests that Oort cloud and Jupiter-family
comets have common origin sites within the early solar 
system \citep[an argument that parallels that derived
from volatile composition studies,][]{ahearn2012}, but the
range of $f_{cryst}$ values in Oort cloud comets suggest this class
may be sampling a particular region that is not represented in the 
Jupiter-family members. This inference is intriguing; however, limited
in robustness as any tentative conclusions are based on a limited
sample size. Large sample sizes are required to substantiate or 
vitiate these trends.

Comet C/2012 K1 (Pan-STARRS) and C/2007 N3 (Lulin) have modest 
mean values for $f_{cryst}$ ($\ltsimeq 48\%$), 
low silicate-to-carbon ratios, and grain size distributions that peak 
at large radii, $\gtsimeq 0.6~\micron$. The 10~\micron{} silicate 
feature is weak and/or absent in these NICs. Perhaps these bodies 
represent a population of more carbon dominated bodies, similar 
to the dark organic KBOs, whose surfaces are devoid of small grains. 
Indeed the low albedo (see \S\ref{sec:c_albedo}) 
of C/2012 K1 (Pan-STARRS) and the dominance of amorphous carbon 
grain materials maybe providing clues.

\subsection{Dust Production Rates}\label{sec:opt_ans}

\begin{figure}
\plotone{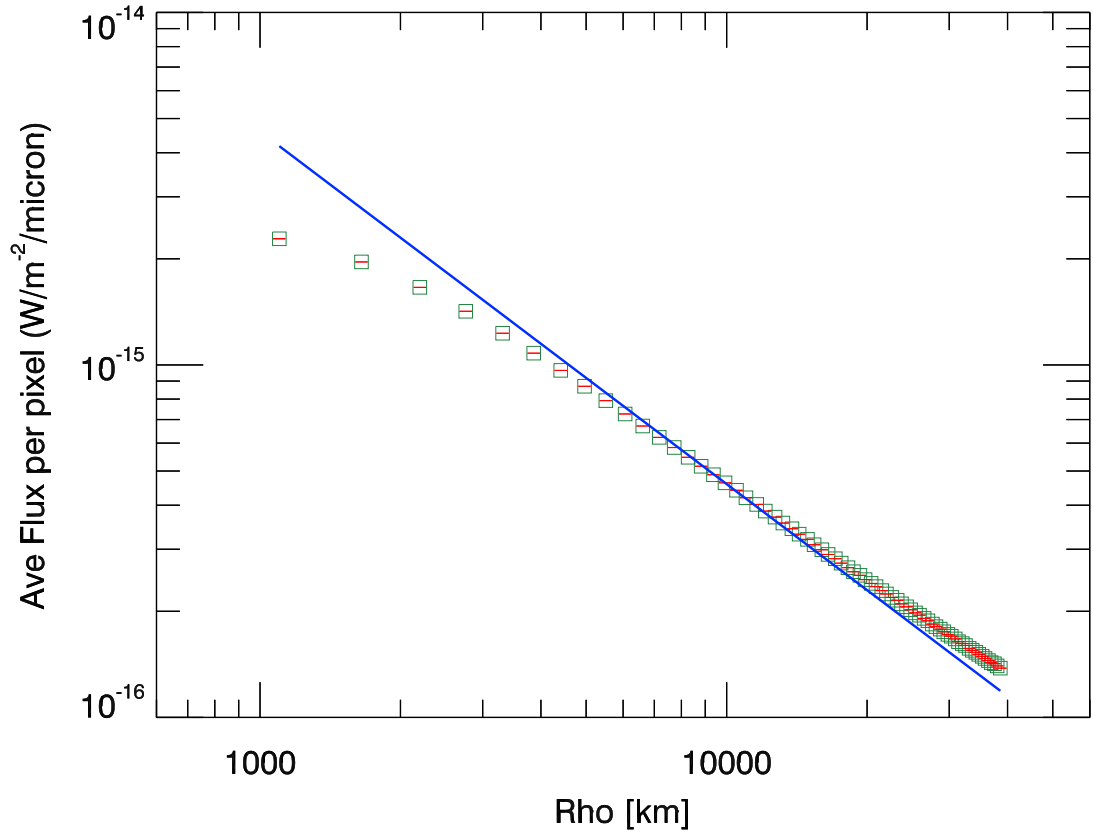}
\caption{Azimuthally averaged radial profile fluxes as a function of linear
radius as measured in the $R$ band from the optical photocenter of
comet C/2012 K1 (Pan-STARRS) obtained on 2014 June 04.24 UT. The solid blue
line denotes a 1/$\rho$ profile.
\label{fig:rradial_profile}}
\end{figure}

The radial profile of comet C/2012 K1 (Pan-STARRS) was plotted to assess the
quality of the data for calculating a dust production rate near
the epoch of our SOFIA observations. The radial
profile of C/2012 K1 (Pan-STARRS) in the $V$ band shows a deviation from the
$1/\rho$ profile \citep{gehrzney92}, suggesting contamination 
from gas such as $C_{2}$ ($\Delta = 0$) band(s) near 5141 \AA. 
Strong $C_{2}$ emission is present in spectra 
\citep[][and also A. McKay, priv. comm.]{mckay2014}
contemporaneous with our optical imagery.
We therefore only calculate the dust production in $R$ band. The 
$R$ band radial profile of C/2012 K1 (Pan-STARRS) is shown in 
Fig.~\ref{fig:rradial_profile}.

To estimate the rate of dust production in comet C/2012 K1 (Pan-STARRS),
we utilize the $Af\rho$ quantity introduced by \citet{ahearn84}. This
quantity serves as a proxy for dust production and when the cometary coma
is in steady state, the value for $A(\Theta)f \rho$ is an aperture independent
parameter,

\begin{equation}
A(\Theta)f \rho = \frac{4  \; r_{h}^{2} \;  \Delta^{2}  \; 10^{-0.4(m_{comet} - m_{\odot})}}{\rho} \; (cm)
\end{equation}

\noindent where $A(\Theta)$ is four times the \textit{geometric} albedo 
at a phase angle $\Theta$, $f$ is the filling factor of 
the coma, $m_{comet}$ is the
measured cometary magnitude, $m_{\odot}$ is the apparent solar 
magnitude, $\rho$ is the linear radius of the aperture
at the comet's position (cm) and $r_{h}$ and $\Delta$ are the
heliocentric and geocentric distances measured in AU and cm, respectively.
To correct our comet measurements for phase angle effects
we applied the Halley-Marcus (HM) \citep{marcus2007a,marcus2007b,schleicher1998}
phase angle correction.\footnote{see \url{http://asteroid.lowell.edu/comet/dustphase.html}}
We adopt an interpolated value of 0.3864 (appropriate for the 
2014 June 04.24~UT dataset) to normalize $A(\Theta)f \rho$ to $0\degr$ 
phase angle. Table~\ref{tab:afr_tab} reports values of 
$Af \rho = [(A({\Theta})f \rho/$HM]
at a selection of distances from the comet photocenter in the $R$-band.

In addition to $Af\rho$, we also compute the $\epsilon f \rho$ parameter of
comet C/2012 K1 (Pan-STARRS) based on our FORCAST broadband photometry
(Table~~\ref{tab:simage_phot_tab}). The $\epsilon f \rho$ parameter 
\citep[defined by][Appendix A]{kelley2013} can be considered 
to be the thermal emission corollary to the scattered-light-based 
$Af\rho$:

\begin{equation}
\epsilon f \rho = \frac{\Delta^2}{\pi \rho} \frac{F_\nu}{B_\nu}\,\rm{(cm)},
\label{eqn:efrhoeqn}
\end{equation}

\noindent where $\epsilon$ is the effective dust emissivity, 
$F_\nu$ is the flux density (Jy) of the comet within the aperture 
$\rho$, $B_\nu$ is the Planck function (Jy/sr) evaluated at the 
temperature $T = T_{scale}\, (278~\rm{K})\, r_{h}^{-0.5}$, where 
the scaling factor $T_{scale} = 0.99$ 
based on the 215~K measured continuum 
temperature discussed in \S\ref{sec:10sef-disc}. Derived 
values of $\epsilon\, f\rho$ for comet C/2012 K1 (Pan-STARRS) 
are presented in Table~\ref{tab:simage_phot_tab}.

\subsection{Coma Averaged Dust Albedo}\label{sec:c_albedo}

Dust albedo is a basic parameter characterizing the size distribution and
physical properties of comet dust that is, surprisingly, infrequently
measured. Following the convention of 
\citet{gehrzney92} the \textit{bolometric} albedo,
($A_{bolometric} \equiv (Energy_{\, scattered} /Energy_{\, incident}$) is

\begin{equation}
A_{bolometric} \simeq \frac{f(\Theta)}{1 + f(\Theta)},
\label{eqn:abgn}
\end{equation} 

\noindent where for comet dust the incident energy is the sum of
the energy scattered by the coma plus the total energy of the
coma's thermal emission at an observed phase angle 
$\Theta$ (Sun-comet-observer angle). The term
$f({\Theta})$ can be determined from fitting the observed spectral
energy distribution of the comet with appropriate Planck blackbody functions
in the infrared (thermal dust emission) and reflected solar spectra 
at optical (scattering) wavelengths

\begin{equation}
f({\Theta}) = \frac{[\lambda{\rm F}_\lambda]_{max, scattering}}{[\lambda{\rm F}_\lambda]_{max, IR}}
\label{eqn:ftheta}
\end{equation}

\noindent where the $[\lambda{\rm F}_{\lambda}]_{max}$ 
is the peak of the SEDs in
the respective wavelength ranges. Lab experiments and theoretical 
calculations of the scattered light from particles indicate that the 
total brightness, color, polarization, and polarization color depend on the
optical constants, particle size distribution, structure, and
porosity of the dust as well as the solar phase
angle \citep{lindsay2013,hadamcik2007,kolok2004}. The spectral shape of the IR 
thermal emission provides a direct link with the mineralogy and grain size.
Both of these processes provide information on the size and composition
of the dust. The scattered light and thermal emission are also
connected to one another through the grain albedo, the ratio of the
scattered light to the total incident radiation.  Because light
is not isotropically scattered by comet dust the measured albedo will
depend not only on the composition and structure of the dust grains, but
also on the phase angle (Sun-comet-observer angle) of the observations.

The coma SED of comet C/2012 K1 (Pan-STARRS) was measured on 2014 June
04~UT using filter photometry at both mid-IR as well as the 
optical (scattered sunlight) wavelengths. These data
enable computation of the coma averaged 
\textit{bolometric} albedo \citep{gehrzney92}. 
Using the integrated flux densities in a circular aperture of
radius $9.989^{\prime\prime}$, $[\lambda{\rm F}_\lambda]_{max, IR}$
was derived from the SOFIA photometry by
$\chi$-square fitting a blackbody to the mid-IR data using
Gaussian weighted errors, resulting in 
T$_{bb} = 214.04 \pm 14.94$~K with a peak flux of 
$ 2.02^{+ 0.12}_{- 0.15} \times 10^{-16}$ W~cm$^{-2}.$
The [V] band photometry is contaminated by gas emission
(\S\ref{sec:opt_ans}). However, the $C_{2}$ bands fall outside
the bandpass of the [R] filter and a 5800~K blackbody 
(the Sun) emission peaks near the [R] filter central 
wavelength ($\lambda_{c} = 0.64$~\micron) in
$\lambda\rm{F}_{\lambda}$ (W~cm$^{-2}$) space. Hence,
$[\lambda{\rm F}_\lambda]_{max, scattering} = 3.33 \pm 0.03 \times 10^{-17}$
W~cm$^{-2}$ derived the [R] band photometry measured in a 
circular aperture of  radius $9.989^{\prime\prime}$
(Table~\ref{tab:afr_tab}).
The dust bolometric albedo of comet 
C/2012 K1 (Pan-STARRS) is $ 0.14 \pm 0.01$ at 
phase angle of $34.76^{\circ}$ (from Eqns.~\ref{eqn:abgn} and \ref{eqn:ftheta}).

\begin{figure}
\plotone{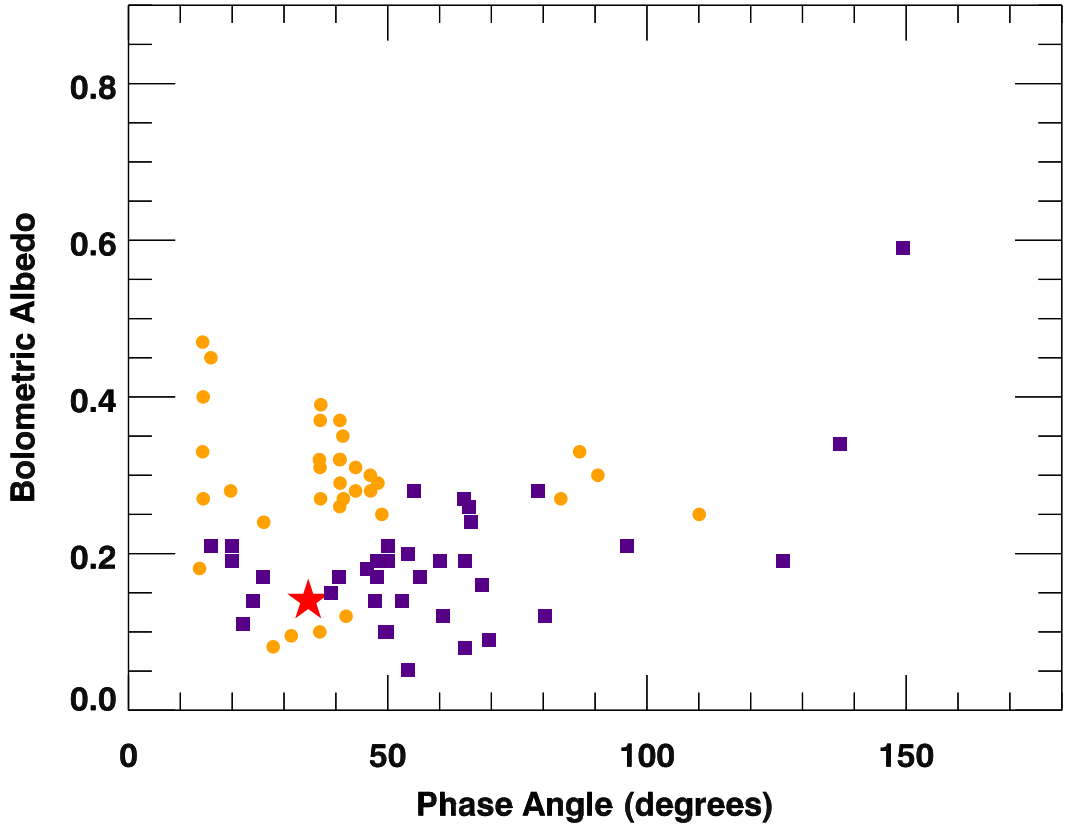}
\caption{The bolometric albedo as a function of phase angle for a
sample of Nearly Isotropic Comets (NICs; filled orange circles) and
Ecliptic Comets (ECs; filled purple squares) derived from the literature
following the prescription of \citep{gehrzney92},
Our measurement of the NIC C/2012 K1 (Pan-STARRS),
$0.14 \pm 0.01$ at a phase angle of
$34.76^{\circ}$, is indicated by the
red star. The phase angles for each comet are obtained from the JPL 
Horizons ephemerides. 
\label{fig:all_comet_albedo}}
\end{figure}

\citet{kolok2004} reviewed published visual albedos of
comets and found only eight 8 comets have measured albedos
\citep[excluding comets Kohoutek and Crommelin discussed in][]{gehrzney92},
and all were from the NIC dynamical class. \citet{mskdw2009}
found only one EC with visual albedo,
21P/Giacobini-Zinner \citep{pittchov2008}. Recently, the albedos of
73P/Schwassmann-Wachmann~3, 103P/Hartley 2, and C/2009 P1 (Garradd)
also have been measured \citep{meech2011,sitko2011,sitko2013}.
Fig.~\ref{fig:all_comet_albedo} is a compilation of the the 
bolometric albedo data that exists on comets,
including our determination for comet C/2012 K1 (Pan-STARRS).
There is considerable scatter for multi-epoch observations of
individual comets. Such scatter arises from variations in activity of a comet
at different epochs of observation. For example, comet C/1995 O1 (Hale-Bopp),
whose data are most scattered, had numerous and fast changing morphological 
structures \citep[jets, shells, envelopes, e.g.,][]{harker1997,woodward1998}. 
All of these features were characterized by differing size and particle 
composition \citep[e.g.,][]{rodriguez1997,schleicher1997}. Thus, the 
difference in the dust albedo for the same comet indicates variations
in comet activity, specifically development of jets and other
morphological features.

The ensemble albedos compiled by \citet{kolok2004} also shows a 
broad distribution of values for each phase angle. The causes for
these latter albedo ranges and the scatter in multi-epoch observations
of comets are unclear, but must reside in the
physical properties of the comet particles, including the grain 
size distribution, porosity, grain structure (i.e., prolate spheroids, crystalline 
needles, etc.), and composition \citep[e.g.,][]{lindsay2013}.  However, 
observations have not yet demonstrated to what extent grain structure or grain 
compositions are important.  To assess these latter aspects, thermal 
emission models and albedo observations of a additional comets are needed.

\section{SUMMARY}
We discuss the pre-perihelion mid-infrared spectrophotometry and
narrow band filter imagery obtained in 2014 June with FORCAST 
on the NASA SOFIA airborne platform of the dynamically 
new comet C/2012 K1 (Pan-STARRS) at a heliocentric distance
of $\simeq 1.70$~AU. The spectral energy distribution of the 
comet at this epoch exhibits a 10~\micron{} silicate feature,
$[F_{10}/F^{BB}_{continuum}] = 1.18 \pm 0.03$ above a blackbody curve 
($T_{bb} = 215.32 \pm 0.95$~K) fit to the spectra continua 
longwards of 12.5~\micron{} which is quite weak compared to comets 
such as C/1999 O1 (Hale-Bopp) or 17P/Holmes. The coma dust bolometric 
albedo, $0.14 \pm 0.01$, derived using contemporaneous optical imagery is 
similar to other comets at the observed phase angle ($\sim 35$\degr), 
while the dust production rate ($Af\rho$) from scattered light observations 
is $\simeq 5340$~cm.

From the observed infrared spectral energy distribution, thermal 
modeling analysis is used to determine the physical characteristics of the coma 
dust population and deduce the silicate crystalline mass 
fraction ($0.20^{ +0.30}_{ -0.10}$)
and silicate-to-carbon dust ratio ($0.80^{ +0.25}_{- 0.20}$). We find 
that grains in the coma of C/2012 K1 (Pan-STARRS) are dominated by
amorphous materials, especially carbon, and the differential grain size
distribution peaks at radii of 0.6~\micron{}, the slope of the 
distribution $N = 3.4$, and the grains are
solid, having a fractal porosity parameter $D = 3.0$. The bulk grain 
properties of comet C/2012 K1 (Pan-STARRS) are comparable to other
Nearly Isotropic comets (NICs) with weak 10~\micron{} silicate features
and similar in respect to coma grains seen in the small-set of 
Ecliptic family comets (ECs) that have fragmented, explosively released 
subsurface materials, or have had materials excavated from depth.

SOFIA observations of comet C/2012 K1 (Pan-STARRS) and other future 
comets enables characterization grain properties in the NIC and EC
dynamical families. These properties, including dust size, porosity, 
and composition, relate to grain formation, radial mixing, and particle 
agglomeration in the proto-solar disk and provide insight to the evolution 
of the early solar system. As the number of well-studied comets increases 
at infrared wavelengths (from whence dust properties can be characterized),
the fundamental differences between comets originating from
different regions and times in the solar system may be eventually discerned.

\section{Acknowledgments}

CEW and his team acknowledge support from Universities Space
Research Association (USRA)/NASA contract NAS2-97001. CEW, MSK, DEH also
acknowledge support from NASA Planetary Astronomy Program grant 
12-PAST12-0016, while CEW and ELR also note support
from NASA Planetary Astronomy Program grant
NNX13AJ11G. The authors would also like to acknowledge the support and 
insight of Drs.~J.~DeBuzier and L.~A. Helton of the SOFIA Science Ctr. for
their assistance with flight planning and data reduction activities.
This work is supported at The Aerospace Corporation by the Independent
Research and Development program. We also thank the comments and suggestions
of an anonymous referee that improved the clarity of our work.

{\it Facilities:} \facility{SOFIA (FORCAST)}, \facility{Bok (90Prime)}

\clearpage

%
%
%

\clearpage

\clearpage

\clearpage

\clearpage

\clearpage

\clearpage

\clearpage


\input{table_1.tex}
\clearpage

\input{table_2.tex}
\clearpage

\input{table_3.tex}
\clearpage

\input{table_4.tex}
\clearpage

\input{table_5.tex}
\clearpage

\end{document}

%% file: table_1.tex

\begin{deluxetable}{lccccccccc}
%
\setlength{\tabcolsep}{2pt}
\tablewidth{0pt}
\tablecaption{SOFIA Observational Summary -- Comet C/2012 K1 (PAN-STARRS)\tablenotemark{*}
\label{tab:sobstab_tab}}
\tablehead{
&&\colhead{Grism} &&\colhead{Total}\\
\colhead{Observation}&&\colhead{or}&&\colhead{On Src}&&&&\colhead{Tail\tablenotemark{a}} &\colhead{Tail\tablenotemark{a}}\\
\colhead{Date} & & \colhead{Fltr} & \colhead{Exp} &\colhead{Integ} &&& \colhead{Phase} &
\colhead{Gas} & \colhead{Dust}\\
\colhead{2014 UT} & \colhead{InstCfg} & \colhead{$\lambda_{c}$} & \colhead{Time} & \colhead{Time} & 
\colhead{$r_{h}$} &\colhead{$\Delta$} &\colhead{Ang} & \colhead{PSAng} & \colhead{PsAMV}\\
\colhead{(dd-mm hr:min:s)} &  & \colhead{($\mu$m)}& \colhead{(sec)}  & \colhead{(sec)}  
& \colhead{(AU)} & \colhead{(AU)} &\colhead{($\circ$)} &\colhead{($\circ$)} & \colhead{($\circ$)}\\
}
\startdata

FOF176 \\
06-04T03:35:31 & Imaging Dual & 19.71 & 29.5 & 616.0 & 1.708 & 1.688 & 34.76 & 105.75 & 78.33\\
06-04T03:35:31 & Imaging Dual & 31.46 & 29.5 & 616.0 \\
06-04T04:16:38 & Imaging Dual & 11.09 & 30.8 & 216.0 \\
06-04T04:16:38 & Imaging Dual & 31.46 & 30.8 & 252.0 \\

\\
FOF177\\
06-06T04:15:14 & Imaging SWC  & 19.71 & 45.0 & 198.0 & 1.684 & 1.711 & 34.76 & 104.16 & 77.47\\
06-06T04:29:07 & Grism LWC & G227 & 22.5 & 1800.0\\               
06-06T04:51:42 & Grism SWC & G111 & 24.0 & 1920.0\\

\\
FOF178\\
06-11T03:53:40 & Imaging SWC  & 19.71 & 42.2 & 204.0 & 1.628 & 1.769 & 34.46 & 100.71 & 75.73\\
06-11T03:57:33 & Grism LWC & G227 & 23.0 & 1748.0\\
06-11T04:32:22 & Grism SWC & G111 & 24.0 & 3336.0 \\
\\
FOF179\\
06-13T04:04:51 & Imaging SWC  & 19.71 & 45.1 & 330.0 & 1.605 & 1.793 & 34.22 & 99.62 & 75.17\\
06-13T04:23:04 & Grism LWC & G227 & 23.0 & 1564.0\\
06-13T04:46:21 & Grism SWC & G111 & 24.0 & 2016.0 \\

\enddata
\tablenotetext{*}{\textbf{Notes.}\, Orbital elements derived from JPL Horizons, ssd.jpl.nasa.gov/horizons.cgi.}
\tablenotetext{a}{Vector direction measured CCW (eastward) from celestial north on the plane of the sky.}

\end{deluxetable}

%% file: table_2.tex
%
%
\begin{deluxetable}{lccccc}
%
\setlength{\tabcolsep}{3pt}
\tablewidth{0pt}
\tablecaption{SOFIA Aperture Photometry and $\epsilon f \rho$ of Comet C/2012 K1 (PAN-STARRS)
\label{tab:simage_phot_tab}}
\tablehead{
\colhead{Observation}\\
\colhead{Date} &  & \colhead{Fltr} & \colhead{Flux} \\
\colhead{UT 2014} & \colhead{InstCfg} & \colhead{$\lambda_{c}$}&\colhead{Density\tablenotemark{a}} &\colhead{$\lambda F_{\lambda}$}&\colhead{$\epsilon f \rho$}\\
\colhead{(dd-mm hr:min:s)}&\colhead{(Imaging)}&\colhead{($\mu$m)}&\colhead{(Jys)}&\colhead{($\times 10^{-16}$ W cm$^{-2}$)}&\colhead{(cm)}\\
}
\startdata

FOF176 \\
06-04T03:35:31 & DUAL & 19.71 & 15.102 $\pm$ 1.117 & 2.297 $\pm$ 0.170 & 14900 $\pm$ 1100 \\
06-04T03:35:31 & DUAL & 31.46 & 12.861 $\pm$ 2.172 & 1.226 $\pm$ 0.207 & 13000 $\pm$ 2200 \\
06-04T04:16:38 & DUAL & 11.09 & 6.119 $\pm$ 1.217  & 1.654 $\pm$ 0.329 & 16300 $\pm$ 3200 \\
06-04T04:16:38 & DUAL & 31.46 & 11.857 $\pm$ 3.149 & 1.130 $\pm$ 0.300 & 11900 $\pm$ 3200 \\

\\
FOF177\\
06-06T04:15:14 & SWC  & 19.71 & 18.338 $\pm$ 2.343 & 2.789 $\pm$ 0.356 & 17900 $\pm$ 2300 \\

\\
FOF178\\
06-11T03:53:40 & SWC  & 19.71 & 16.962 $\pm$ 2.246 & 2.580 $\pm$ 0.342 & 16100 $\pm$ 2100 \\

\\
FOF179\\
06-13T04:04:51 & SWC  & 19.71 & 16.991 $\pm$ 1.777 & 2.584 $\pm$ 0.270 & 16000 $\pm$ 1700 \\

\enddata
\tablenotetext{a}{Measured in a circular aperture with a radius of 9.984$^{\prime\prime}$ 
centroided on the photocenter of the comet nucleus.}

\end{deluxetable}

%% file: table_3.tex

\begin{deluxetable}{lcc}
\tablewidth{0pt}
\tablecaption{Best-fit Thermal Model Parameters and Derived Grain Mineralogy of 
Comet C/2012 K1 (Pan-STARRS)\tablenotemark{*}
\label{tab:bfmods_tab}}
\tablehead{
           &                        & Sub-\micron{} \\
  Dust component & $N_{p}$\tablenotemark{a} $ \times 10^{16}$ & mass fraction
}
\startdata
Amorphous pyroxene    &$8526.433^{+  1067.813}_{-  1494.938}$ &$0.310^{+ 0.043}_{-  0.060}$ \\
\\
Amorphous olivine     &$1228.326^{+   213.563}_{-   640.688}$ &$0.045^{+ 0.009}_{-  0.026}$ \\
\\
Amorphous carbon      &$15246.560^{+   213.563}_{-   427.125}$ &$0.555^{+ 0.009}_{-  0.017}$ \\
\\
Crystalline olivine   &$2476.880^{+  2135.626}_{-   854.250}$  &$0.090^{+ 0.078}_{-  0.031}$ \\
\\
Crystalline pyroxene  &$0.000^{+  1067.813}_{-     0.000}$     &$0.000^{+ 0.043}_{- 0.000}$ \\
\cutinhead{Other model parameters}
$\chi^2_\nu$            & 0.98 \\
\\
Degrees of freedom    & 156 \\
\\
Total submicron grain mass\tablenotemark{b}  & $(7.663^{+ 1.310}_{- 0.952}) \times 10^{5}$ kg \\
\\
Silicate/carbon ratio     & $0.80^{+  0.25}_{-  0.20}$ \\
\\
$f_{cryst}$           & $0.202^{+ 0.297}_{- 0.099}$\\

\enddata
\tablenotetext{*}{\textbf{Notes.}\, Uncertainties represent the 95\% confidence level.}
\tablenotetext{a}{Number of grains at the peak of the grain size distribution.}
\tablenotetext{b}{The total mass of the sub-\micron{} sized grains contained within the 
spectral extraction aperture.}
\end{deluxetable}

%% file: table_4.tex

\begin{deluxetable}{lcccccc}
%
\setlength{\tabcolsep}{2pt}
\tablewidth{0pt}
\tablecaption{Thermal Modeling Dust Characteristics of Select Comets
\label{tab:fcryst_tab}}
\tablehead{
&&\colhead{$f_{cryst}$\tablenotemark{a}} &\colhead{SAC\tablenotemark{b}}\\
\colhead{Comet Class} &\colhead{1/$a_{orig}$} &\colhead{Range} &\colhead{Ratio} &\colhead{$N$} &\colhead{$a_{peak}$}  &\colhead{Refs\tablenotemark{*}}\\
\colhead{} &\colhead{($10^{-6}$ AU$^{-1}$)} &\colhead{(\%)} &&&\colhead{($\mu$m)}
}
\startdata

\underline{NIC/OC\tablenotemark{c}}\\
\\
C/2012 K1 (Pan-STARRS) & 42.9 & 10-50 & 0.6-1.1    & 3.4 & 0.6 & This work \\

C/2007 N3 (Lulin)     & 32.2 & 34-51  & 0.42-0.54    & 4.2 & 0.9 & [1] \\

C/2001 Q4 (NEAT)      & 61.2 & 71  & 1.7-5.7 & 3.7 & 0.3 & [2,3] \\


C/2002 V1 (NEAT)          & 2279.3 & 66-69  & 1.38    & 3.7  & 0.5 & [3] \\

C/1995 O1 (Hale-Bopp)     & 3805.0 & 60-78   & 8.1-13.3    & 3.4-3.7  & 0.2 & [4,5] \\

\\
\underline{EC/JFC\tablenotemark{c}}\\
\\

9P/Tempel 1 ($\sim1$hrs post-impact, ctr) & $\ldots$ & 19-25 & 3.4-4.4 & 3.7 & 0.2 & [7] \\

17P/Holmes  & $\ldots$ & $\sim42$\tablenotemark{c}  & 0.2 & $\ldots$  & $\ldots$ & [7] \\

73P/SW3-B (Apert B) & $\ldots$ & 43-69 & 1.09-1.59 & 3.4  & 0.5 & [8] \\

73P/SW3-C  (Apert M) & $\ldots$ & 57-69 & 0.60-0.75 & 3.4 & 0.3 & [8] \\

\enddata
\tablenotetext{a}{Computed from thermal model sub-\micron{} silicate mass 
fractions using Eqn.(\ref{eqn:fcseqn}) described in \S\ref{sec:fcryst-disc}, where 
the range includes the model uncertainties when known.}
\tablenotetext{b}{Defined as the silicate-to-amorphous carbon ratio derived from
thermal modelling of the SEDs where the range 
includes the model uncertainties when known.}
\tablenotetext{c}{Comet dynamical class divisions are NIC/OC = Nearly Isotropic/Oort Cloud; 
EC/JFC = Ecliptic/Jupiter-Family}
\tablenotetext{d}{Estimated from \citet{reach2010} who provide abundances weighted 
by grain surface area.}
\tablenotetext{*}{\textbf{References.}\, (1) \citet{cew2011}; (2) \citet{wooden2004}; 
(3) \citet{ootsubo2007}; (4) \cite{harker2002}; (5) \cite{harker2004a};
(6) \cite{harker2007}; (7) \cite{reach2010}; (8)\cite{harker2011}}

\end{deluxetable}

%% file: table_5.tex

\begin{deluxetable}{cccc}
\tablewidth{0pt}
\tablecaption{Af$\rho$ Values for Comet C/2012 K1 (PanSTARRS) on 2014 June 04.24~UT
\label{tab:afr_tab}}
\tablehead{\colhead{Aperture \tablenotemark{a}} & \colhead{$\rho$} 
& \colhead{$R$} & \colhead{Af$\rho$ \tablenotemark{b}}\\
\colhead{(arcsec)} & \colhead{(km)} & \colhead{(mag)} & \colhead{(cm)}}
\startdata
11.24 & 6705  & 12.157$\pm$0.008 & 5731 $\pm$ 42\\
19.97 & 11912 & 11.593$\pm$0.009 & 5424 $\pm$ 43\\ 
25.14 & 14999 & 11.344$\pm$0.009 & 5417 $\pm$ 43\\ 
33.53 & 19999 & 11.065$\pm$0.009 & 5253 $\pm$ 41\\ 
41.91 & 24999 & 10.835$\pm$0.009 & 5194 $\pm$ 41\\ 
50.29 & 29999 & 10.674$\pm$0.009 & 5020 $\pm$ 39\\ 
\enddata
\tablenotetext{a}{Effective circular aperture diameter.}
\tablenotetext{b}{Af$\rho$ values corrected to zero phase (see \S\ref{sec:opt_ans}).}
\end{deluxetable}

%% file: archiveX_ms.bbl
\begin{thebibliography}{}

\bibitem[A'Hearn et al.(2012)]{ahearn2012} A'Hearn, M.~F., Feaga, 
L.~M., Keller, H.~U., et al.\ 2012, \apj, 758, 29 

\bibitem[A'Hearn et al.(1995)]{ahearn1995} A'Hearn, M.~F., Millis, R.~L.,
Schleicher, D.~G., Osip, D.~J., \& Birch, P.V.\ 1995, \icarus, 118, 223

\bibitem[A'Hearn et al.(1984)]{ahearn84} A'Hearn, M.~F., Schleicher, D.~G.,
Feldman, P.~D., Millis, R.~L., \& Thompson, D.~T. 1984, \aj, 89, 579

\bibitem[Berger et al.(2011)]{berger2011} Berger, E.~L., Zega, 
T.~J., Keller, L.~P., \& Lauretta, D.~S.\ 2011, \gca, 75, 3501 

\bibitem[Bockel{\'e}e-Morvan et al.(2002)]{bv2002} Bockel{\'e}e-Morvan, D., Gautier, 
D., Hersant, F., Hur{\'e}, J.-M., \& Robert, F.\ 2002, \aap, 384, 1107 

\bibitem[Bradley \& Dai(2004)]{bradleydai2004} Bradley, J.~P., \& Dai,
Z.~R.\ 2004, \apj, 617, 650

\bibitem[Bradley \& Dai(2000)]{bradleydai2000} Bradley, J.~P.,
\& Dai, Z.\ 2000, Meteoritics and Planetary Science Supplement, 35, 32

\bibitem[Bradley et al.(1999)]{bradley1999} Bradley, J.~P., Keller, 
L.~P., Gezo, J., et al.\ 1999, Lunar and Planetary Science Conference, 30, 
1835 

\bibitem[Brownlee(2014)]{brownlee2014} Brownlee, D.\ 2014, Ann. Rev. Earth 
Planet. Sci. 42, 179

\bibitem[Brownlee et al.(2012)]{brownlee2012} Brownlee, D., Joswiak, D.,
Matrajt, G.\ 2012, Meteoritics and Planetary Science, 47, 453

\bibitem[Brownlee et al.(2006)]{brownlee2006} Brownlee, D., Tsou, P., Al{\'e}on,
et al.\ 2006, Science, 314, 1711

\bibitem[Bursentova et al.(2012)]{bursentova2012} Brusentsova, T., 
Peale, R.~E., Maukonen, D., et al.\ 2012, \mnras, 420, 2569 

\bibitem[Ciesla \& Sandford(2012)]{cieslasanford2012} Ciesla, F.~J., 
Sandford, S.~A.\ 2012 Science, 336, 452

\bibitem[Cielsa(2011)]{ciesla2011} Ciesla, F.~J.\ 2011, \apj, 740, 9

\bibitem[Ciesla(2007)]{ciesla2007} Ciesla, F.~J.\ 2007, Science, 318, 613 

\bibitem[Charnoz \& Morbidelli(2007)]{cm2007} Charnoz, S., \&  
Morbidelli, A.\ 2007, Icarus, 188, 468

\bibitem[Clarke, Vacca, \& Shuping(2014)]{clarke2014} Clarke, M., Vacca, W.~D.,
\& Shuping, R.~Y.\ 2014 i nADASS Conf, Ser., ADASS XXIV, eds. A.~R. Taylor
\& J.~M. Stil [San Francisco, CA: ASP]

\bibitem[Davoisne et al.(2006)]{davoisne2006} Davoisne, G., Djouadi, Z., 
Leroux, H., et al.\ 2006, \aap, 448, L1

\bibitem[Dones et al.(2004)]{dones2004} Dones, L., Weissman, P.~R., 
Levison, H.~F.  \& Duncan, M.~J.\ 2004, in Comets II, eds. 
M~C~. Festou, H.~U. Keller, and H.~A. Weaver, [University of Arizona 
press: Tucson AZ], p.153ff

\bibitem[Fabian et al.(2000)]{fabian2000} Fabian, D., Jager, C., 
Henning, Th., et al.\ 2000, \aap,364, 282 

\bibitem[Formenkova(1999)]{formenkova1999} Formenkova, M.~N.\ 1999, Space 
Sci. Rev. 90, 109

\bibitem[Fomenkova et al.(1994)]{formenkova1994} Fomenkova, M.~N., 
Chang, S., \& Mukhin, L.~M.\ 1994, \gca, 58, 4503 

\bibitem[Flynn et al.(2013)]{flynn2013} Flynn, G.~J., Wirick, S., Keller, 
L.~P., et al.\ 2013, Earth, Planets, and Space, 65, 1159

\bibitem[Flynn et al.(2003)]{flynn2003} Flynn, G.~J., Keller, 
L.~P., Feser, M., Wirick, S., \& Jacobsen, C.\ 2003, \gca, 67, 4791 

\bibitem[Gehrz et al.(2009)]{gehrz2009} Gehrz, R.~D., Becklin, E.~E., 
de Pater, I., Lester, D.~F., Roellig, T.~L., \& Woodward, C.~E.\ 2009, 
AdSpR 44, 413

\bibitem[Gehrz et al.(1995)]{gehrz1995} Gehrz, R.~D., Johnson, C.~H., 
Magnuson, S.~D., \& Ney, E.~P.\ 1995, \icarus, 113, 129

\bibitem[Gehrz \& Ney(1992)]{gehrzney92} Gehrz, R.~D., \& Ney, E.~P.\ 1992,
\icarus, 100, 162

\bibitem[Gicquel et al.(2012)]{gicquel2012} Gicquel, A., Bockel{\'e}e-Morvan,
D., Zakharov, V. V., Kelley, M. S., Woodward, C. E., \&
Wooden, D. H.\ 2012, \aap, 542, 119

\bibitem[Gomes et al.(2005)]{gomes2005} Gomes, R., et al.\ 2005, 
Nature 435, 466

\bibitem[Hadamcik et al.(2007)]{hadamcik2007} Hadamcik, E., et al.\ 2007, 
\icarus, 190, 459

\bibitem[Hanner et al.(1994)]{hanner1994} Hanner, M.~S., Lynch,  D.~K.,
\& Russell, R.~W.\ 1994, \apj, 425, 274

\bibitem[Hanner \& Zolensky(2010)]{hannerzol2010} Hanner, M.~H., \& 
Zolensky, M.~E.\ 2010,
Lecture Notes in Physics, Berlin Springer Verlag, 815, 203

\bibitem[Harker \& Desch(2002)]{dehsjd2002} Harker, D. E., \&
Desch, S. J.\ 2002, \apj, 565, L109

\bibitem[Harker et al.(1997)]{harker1997} Harker, D.~E., Woodward, C.~E., 
McMurtry, C.~W., et al.\ 1997, Earth Moon and Planets, 78, 259

\bibitem[Harker et al.(2002)]{harker2002} Harker, D.~E., Wooden, D.~H., 
Woodward, C.~E., \& Lisse, C.~M.\ 2002, \apj, 580, 579

\bibitem[Harker et al.(2004a)]{harker2004a} --. 2004a, Erratum:  \apj, 615, 1081

\bibitem[Harker et al.(2004b)]{harker2004b} Harker, D. E., Woodward, C. E.,
Wooden, D. H., \& Kelley, M. S.\ 2004b, AAS, 205, 5612H

\bibitem[Harker et al.(2005)]{harker2005} Harker, D. E., Woodward, C. E., 
\& Wooden, D. H.\ 2005, Science, 310, 278

\bibitem[Harker et al.(2007)]{harker2007} Harker, D. E., Woodward, C. E., 
Wooden, D. H., et al.\ 2007, \icarus, 190, 432

\bibitem[Harker et al.(2011)]{harker2011} Harker, D. E., Woodward, C. E., 
Kelley, M. S., Sitko, M. L., Wooden, D. H., Lynch, D. K., \&  
Russell, R. W.\ 2011, \aj, 141, 26

\bibitem[Heck et al.(2012)]{heck2012} Heck, P.~R., Hoppe, P., \&
Huth, J.\ 2012, Meteoritics and Planetary Science, 47, 649

\bibitem[Henning(2003)]{henning2003} Henning, T.\ 2003, in Lecture
Notes in Physics, Vol. 609, Astromineralogy, eds. T. K.
Henning, (Springer-Verlag: Berlin), pp.266

\bibitem[Henning(2010)]{henning2010} Henning, Th.\ 2010, \araa, 48, 21

\bibitem[Herter et al.(2012)]{herter2012} Herter, T.~L., Adams, J.~D., \&
de Buizer, J.~M.\ 2012, \apj, 749, L18

\bibitem[Hill(2001)]{hill2001} Hill, P.~M. et al.\ 2001, Pub.\ Nat.\
Acad.\ Sci.\ 91, No.5, 2182

\bibitem[H{\"o}rz et al.(2006)]{horz2006} H{\"o}rz, F., Bastien,
R., Borg, J., et al.\ 2006, Science, 314, 1716

\bibitem[Hony et al.(2002)]{hony2002} Hony, S., Bouwman, J., Keller, L.~P., 
\& Waters, L.~B.~F.~M.\ 2002, \aap, 393, L103

\bibitem[Hughes \& Armitage(2010)]{ha2010} Hughes, A. L. H.,
\& Armitage, P. J.\ 2010, \apj, 719, 1633

\bibitem[Ishii et al.(2008)]{ishii2008} Ishii, H.~A., et al.\ 2008, 
Science, 319, 447

\bibitem[Jewitt(2007)]{jewitt2007} Jewitt, D. 2007, in Trans-Neptunian
Objects and Comets, Saas-Fee Advanced Course 35, v35, p1
[Springer-Verlag: Berlin]

\bibitem[Joswiak et al.(2012)]{joswiak2012} Joswiak, D.~J., 
Brownlee, D.~E., Matrajt, G., et al.\ 2012, Lunar and Planetary Science 
Conference, 43, 2395 

\bibitem[Laher et al.(2012)]{laher2012} Laher, R.~R., et al.\ 2012,
\pasp, 124, 737

\bibitem[Li et al.(2015)]{lijy2015} Li, J.-Y., Thomas, P.~C.,
Veverka, J., et al.\ 2015, Highlights of Astronomy, 16, 180

\bibitem[Lindsay et al.(2013)]{lindsay2013} Lindsay, S.~S., Wooden,
D.~H., Harker, D.~E., et al.\ 2013, \apj, 766, 54

\bibitem[Lisse et al.(2006)]{lisse2006} Lisse, C. M., et al.\ 2006,
Science, 313, 635

\bibitem[Kelley \& Wooden(2009)]{mskdw2009} Kelley, M.~S., 
\& Wooden, D.~H.\ 2009, Planet. Space Sci., 57, 1133

\bibitem[Kelley et al.(2015)]{kelley2015} Kelley, M.~S., Woodward, C.~E.,
Harker, D.~E., Wooden, D.~H, Sitko, M.~L., Russel, R.~W., \& 
Kim, D.~L.\ 2015a, AAS, 2254, 305K

\bibitem[Kelley et al.(2013)]{kelley2013} Kelley, M.~S., Fern{\'a}ndez, Y.~R., 
Licandro, J., et al.\ 2013 \icarus, 225, 475

\bibitem[Kelley et al.(2006)]{kelley2006} Kelley, M.~S., et al.\ 2006,
\apj, 651, 1256

\bibitem[Keller et al.(2007)]{keller2007} Keller, H.~U., 
K{\"u}ppers, M., Fornasier, S., et al.\ 2007, \icarus, 191, 241 

\bibitem[Keller et al.(2002)]{keller2002} Keller, L.~P., Hony, S., 
Bradley, J.~P., et al.\ 2002, \nat, 417, 148 

\bibitem[Kemper et al.(2004)]{kemper2004} Kemper, F., Vriend, W.~J., \&
Tielens, A.~G.~G.~M.\ 2004, \apj, 609, 826

\bibitem[Kemper et al.(2005)]{kemper2005} --. 2005, Erratum:, \apj, 633, 534

\bibitem[Kobayashi et al.(2013)]{kobayashi2013} Kobayashi, H., Kimura, H., \&
Yamamoto, S.\ 2013, aap, 550, 72

\bibitem[Koike et al.(2010)]{koike2010} Koike, C., et al.\ 2010, \apj, 709, 983

\bibitem[Kolokolova et al.(2004)]{kolok2004} Kolokolova, L., et al.\ 2004, 
in Comets II, eds.  M.~C. Festou, H.~U. Keller, \& H.~A. Weaver, 
[U. of Arizona Press, Tucson], p.577

\bibitem[Levison(1996)]{levison1996} Levison, H. F.\ 1996, Comet Taxonomy, in
ASPC, Vol. 107, ed. T. Rettig \& J. M. Hahn, (ASP: Tucson), pp.173-191

\bibitem[Levison(2006)]{levison2006} Levison, H.~F., et al.\ 2006, 
\icarus, 184, 619

\bibitem[Levison(2009)]{levison2009} Levison, H.~F., et al.\ 2009, 
Nature, 260, 364

\bibitem[Li \& Draine(2001)]{lidraine2001} Li, A., \& Draine, B. T.\ 2001,
\apj, 550, L213

\bibitem[Li \& Greenberg(1998)]{ligrb1998} Li, A., \& Greenberg, 
J.~M.\ 1998, \aap, 338, 364

\bibitem[Marcus(2007a)]{marcus2007a} Marcus, X.\ 2007a, International
Comet Qrtly April, 39

\bibitem[Marcus(2007b)]{marcus2007b} Marcus, X.\ 2007b, International
Comet Qrtly October, 119

\bibitem[Matrajt et al.(2008)]{matrajt2008} Matrajt, G., Ito, M., Wirick, S., 
et al.\ 2008, Meteoritics and Planetary Science, 43, 315

\bibitem[Matsuno et al.(2012)]{matsuno2012} Matsuno, J., 
Tsuchiyama, A., Koike, C., et al.\ 2012, \apj, 753, 141 

\bibitem[McKay et al.(2014)]{mckay2014} McKay, A., Kelley, M., Cochran, A.,
Dello Russo, N., DiSanti, M., Lisee, C., \& Chanover, N.\ 2014,
DPS, 46, 11002

\bibitem[Meech et al.(2011)]{meech2011} Meech, K., et al.\ 2011, \apj, 734, L1

\bibitem[Min et al.(2005)]{min2005} Min, M., Hovenier, J.~W., 
de Koter, A., Waters, L.~B.~F.~M., \& Dominik, C.\ 2005, \icarus, 179, 158

\bibitem[Nakamura-Messenger et al.(2011)]{nakamura2011} Nakamura-Messenger, K., 
Clemett, S.~J., Messenger, S., \& Keller, L.~P.\ 2011, Meteoritics and 
Planetary Science, 46, 843 

\bibitem[Ogliore et al.(2011)]{ogliore2011} Ogliore, R.~C., et
al.\ 2011, \apj, 745, L19

\bibitem[Olofssson et al.(2010)]{olofsson2010} Olofsson, J., et 
al.\ 2010, \aap,, 520, A39

\bibitem[Oort(1950)]{oort1950} Oort, J.~H.\ 1950, \bain, 11, 91

\bibitem[Ootsubo et al.(2007)]{ootsubo2007} Ootsubo, T., Watanabe, J.-I., 
Kawakita, H., Honda, M., \& Furusho, R.\ 2007, \planss, 55, 1044 

\bibitem[Pittichov{\'a} et al.(2008)]{pittchov2008} Pitticov{\'a} J., 
Woodward, C.~E., Kelley, M.~S., \& Reach, W.~T.\ 2008, \aj, 136, 112`

\bibitem[Reach et al.(2010)]{reach2010} Reach, W.~T., Vaubaillon, J., 
Lisse, C.~M., Holloway, M., \& Rho, J.\ 2010,Icarus, 208, 276

\bibitem[Rodriguez et al.(1997)]{rodriguez1997} Rodriguez, E., Ortiz, J.~L., 
Lopez-Gonzalez, M.~J., et al.\ 1997, \aap, 324, L61 

\bibitem[Sandford et al.(2006)]{sandford2006} Sandford, S.~A., 
Al{\'e}on, J., Alexander, C.~M.~O., et al.\ 2006, Science, 314, 1720 

\bibitem[Schleicher et al.(1997)]{schleicher1997} Schleicher, D.~G., 
Lederer, S.~M., Millis, R.~L., \& Farnham, T.~L.\ 1997, Science, 275, 1913

\bibitem[Schleicher et al.(1998)]{schleicher1998} Schleicher, D.~G., et 
al.\ 1998, \icarus, 132, 397

\bibitem[Schulz et al.(2015)]{schulz2015} Schulz, R., Hilchenbach, 
M., Langevin, Y., et al.\ 2015, \nat, 518, 216 

\bibitem[Sitko et al.(2013)]{sitko2013} Sitko, M.~L., Russell, R.~W., 
Woodward, C.~E., et al.\ 2013, Lunar and Planetary Science Conference, 44, 1154 

\bibitem[Sitko et al.(2011)]{sitko2011} Sitko, M.~L., Lisse, C.~M., 
Kelley, M.~S., et al.\ 2011, \aj, 142, 80 

\bibitem[Sitko et al.(2004)]{sitko2004} Sitko, M.~L., Lynch, D.~L., 
Russell, R.~W., \& Hanner, M.~S.\ 2004, \apj, 612, 576

\bibitem[Stodolna et al.(2012)]{stodolna2012} Stodolna, J., Jacob, 
D., \& Leroux, H.\ 2012, \gca, 87, 35 

\bibitem[Sugita et al.(2005)]{sugita2005} Sugita, S., Ootsubi, T., Kadono, T.,
et al.\ 2005, Science, 310, 274

\bibitem[Temi et al.(2014)]{temi2014} Temi, P., Marcum, P.~M., Young, E., et al.\
2014, \apjs, 212, 24

\bibitem[Velbel \& Harvey(2007)]{velbel2007} Velbel, M.~A., \& Harvey, R.~P.\ 2007, 
Lunar and Planetary Science Conference, 38, 1700

\bibitem[Walsh et al.(2011)]{walsh2011} Walsh, K.~J., et al.\ 2011, Nature, 475, 206

\bibitem[Watanabe et al.(2009)]{watanabe2009} Watanabe, J.-I., 
Honda, M., Ishiguro, M., et al.\ 2009, \pasj, 61, 679 

\bibitem[Watson et al.(2009)]{watson2009} Watson, D.~M., Leisenring, J.~M.,
Furlan, E., et al. \ 2009, \apjs, 180, 84

\bibitem[Wehrstedt \& Gail(2008)]{wg2008} Wehrstedt, M., \& Gail, H.-P.\ 2008, arXiv:0804.3377

\bibitem[Williams et al.(2004)]{williams04} Williams, G.~G., Olszewski,
E., Lesser, M.~P., \&  Burge, J.~H. 2004, \procspie, 5492, 787

\bibitem[Williams(2015)]{williams2015} Williams, G.~V.\ 2015, Observations and
Orbits of Comets, Minor Planet Elec. Circ., 2015-A10

\bibitem[Wirick et al.(2009)]{wirick2009} Wirick, S., Flynn, G.~J., 
Keller, L.~P., et al.\ 2009, Meteoritics and Planetary Science, 44, 1611 

\bibitem[Wooden(2008)]{wooden2008} Wooden, D.~H.\ 2008, Space Sci. Rev., 138, 75

\bibitem[Wooden et al.(1999)]{wooden1999} Wooden, D.~H., Harker, D.~E.,
Woodward, C.~E., et al.\ 1999, \apj, 517, 1034

\bibitem[Wooden et al.(2004)]{wooden2004} Wooden, D.~H., Woodward, 
C.~E., \& Harker, D.~E.\ 2004, \apjl, 612, L77 

\bibitem[Wooden et al.(2005)]{wooden2005} Wooden, D.~H., Harker, D.~E., 
\& Brearley, A.~J.\ 2005, ASP Conf. Series, v341, eds. 
A.~N. Krot, E.~R.~D. Scott, \& B. Reipurth. (San Francisco: ASAP), p.774

\bibitem[Wooden et al.(2011)]{wooden2011} Wooden, D.~H., Woodward, C.~E.,
Kelley, M.~S., Harker, D.~E., et al.\ 2011, EPSC Abs. Vol.~6, EPSC-DPS2011-1557

\bibitem[Woodward et al.(1998)]{woodward1998} Woodward, C.~E., Gehrz, R.~D., 
Mason, C.~G., Jones, T.~J., \& Williams, D.~M.\ 1998, Earth Moon and Planets, 81, 217 

\bibitem[Woodward et al.(2007)]{cew2007} Woodward, C.~E., Kelley, M.~S.,
Bockel{\'e}e-Morvan, D., \& Gehrz, R.~D.\ 2007, \apj, 671, 1065

\bibitem[Woodward et al.(2011)]{cew2011} Woodward, C.~E., et al.\ 2011, 
\aj, 141, 181

\bibitem[Woodward et al.(2013)]{woodward2013} Woodward, C.~E., Russell,
R.~W., Harker, D.~E., Kim, D.~L., Cabreira, B., Sitko, M.~L., Wooden,
D.~H., \& Kelley, M.~S.\ 2013, IAUC 9256

\bibitem[Young et al.(2012)]{young2012} Young, E.~T., Becklin, E.~E.,
Marcum, P.~M., Roellig, T.~L., et al.\ 2012, \apj, 749, L17

\bibitem[Zacharias et al.(2004)]{zacharias2004} Zacharias, N., Monet, D.~G., 
Levine, S.~E., Urban, S.~E., Gaume, R., Wycoff, G.~L.\ 2004, 
Bulletin of the American Astron. Soc. 36, 1418 

\bibitem[Zolensky et al.(2008)]{zolensky2008} Zolensky, M.~E., 
Nakamura-Messenger, K., Rietmeijer, F., et al.\ 2008, Meteoritics and 
Planetary Science, 43, 261

\bibitem[Zolensky et al.(2006)]{zolensky2006} Zolensky, M.~E., Zega, 
T.~J., Yano, H., et al.\ 2006, Science, 314, 1735


\end{thebibliography}
